\documentclass[journal]{IEEEtran}
\IEEEoverridecommandlockouts
\usepackage{cite}
\usepackage[colorlinks=true,linkcolor=black,anchorcolor=black,citecolor=black,filecolor=black,menucolor=black,runcolor=black,urlcolor=magenta]{hyperref}

\usepackage{amsmath,amssymb,amsfonts,mathrsfs}
\usepackage{algorithmic}
\usepackage{graphicx}
\usepackage{textcomp}
\usepackage{xcolor}
\usepackage{tabularx,booktabs}
\newcolumntype{C}{>{\centering\arraybackslash}X} 
\def\BibTeX{{\rm B\kern-.05em{\sc i\kern-.025em b}\kern-.08em
    T\kern-.1667em\lower.7ex\hbox{E}\kern-.125emX}}

\ifCLASSOPTIONcompsoc
\usepackage[caption=false,font=normalsize,labelfont=sf,textfont=sf]{subfig}
\else
\usepackage[caption=false,font=footnotesize]{subfig}
\fi

\newcommand{\bs}[1]{\boldsymbol{#1}}

\begin{document}
\bstctlcite{Settings}
\title{Symmetric--Reciprocal--Match Method for\\ Vector Network Analyzer Calibration}

\author{%
	\IEEEauthorblockN{%
	Ziad~Hatab, Michael~Ernst~Gadringer, and~Wolfgang~Bösch%\\
	%Graz University of Technology, Austria\\
	%\{z.hatab; michael.gadringer; wbosch\}@tugraz.at
	}%
	\thanks{This work was supported by the Christian Doppler Research Association and by the Austrian Federal Ministry for Digital and Economic Affairs and the National Foundation for Research, Technology, and Development}
	\thanks{Ziad~Hatab, Michael~Ernst~Gadringer and Wolfgang~Bösch are with the Institute of Microwave and Photonic Engineering, Graz University of Technology, 8010 Graz, Austria, and also with the Christian Doppler Laboratory for Technology Guided Electronic Component Design and Characterization (TONI), 8010 Graz, Austria (e-mail: \{z.hatab, michael.gadringer, wbosch\}@tugraz.at).}
	\thanks{Software implementation and measurements are available online:\\
	\url{https://github.com/ZiadHatab/srm-calibration}}
}%
\markboth{This work has been accepted for publication in the IEEE Transactions on Instrumentation and Measurement}{}
\maketitle

\begin{abstract}
This paper proposes a new approach, the symmetric-reciprocal-match (SRM) method, for calibrating vector network analyzers (VNAs). The method involves using multiple symmetric one-port loads, a two-port reciprocal device, and a matched load. The load standards consist of two-port symmetric one-port devices, and at least three unique loads are used. However, the specific impedances of the loads are not specified. The reciprocal device can be any transmissive device, although a non-reciprocal device can also be used if only the one-port error boxes are of interest. The matched load is fully defined and can be asymmetric. We demonstrated the accuracy of the proposed method with measurements of coaxial standards using a commercial METAS traceable short-open-load-reciprocal (SOLR) calibration kit with verification standards. In addition, we presented a numerical Monte Carlo analysis considering various uncertainty factors. An advantage of the proposed method is that only the match standard is defined, whereas the remaining standards are partially defined, either through symmetry or reciprocity.
\end{abstract}


\section{Introduction}
\label{sec:1}

\IEEEPARstart{T}{he} most commonly used method for calibrating vector network analyzers (VNAs) is the short-open-load-thru (SOLT) method \cite{Kruppa1971}, which requires that all four standards to be fully characterized or modeled. In the past, many VNAs used a three-sampler architecture with three receivers.To account for the non-driving port's termination mismatches (switch terms), the VNA is modeled with the well-known 12-term model \cite{Rumiantsev2008}. This model forms the foundation of the SOLT calibration.

Nowadays, modern VNAs use a full-reflectometry architecture that allows for sampling all waves, thus directly measuring the switch terms of a VNA by simply connecting a transmissive device between the ports \cite{Jargon2018}. This upgraded architecture enabled the use of the simplified error box model of VNAs \cite{Marks1997}, which has led to many new advanced calibration methods that surpass the accuracy of SOLT \cite{Rumiantsev2008}. Furthermore, even with the three-sampler VNA architecture, it is possible to indirectly measure the switch terms of the VNA using a set of reciprocal devices, which enable the application of the error box model \cite{Hatab2023c}. A well-known family of calibration methods based on the error box model is the self-calibration methods \cite{Rumiantsev2008}, which do not require full characterization of some of the standards. One of the most used self-calibration methods nowadays is the short-open-load-reciprocal (SOLR) method \cite{Ferrero1992}, which is the same as SOLT, but with any transmissive reciprocal device instead of the thru standard. SOLR has proven useful in scenarios where a direct connection is unavailable. However, the drawback of the SOLR method is the requirement of the full definition of the short-open-load (SOL) standards, which bounds the accuracy of SOLR to the SOL standards.

Other self-calibration methods include thru-reflect-line (TRL) and multiline TRL \cite{Engen1979,Marks1991,Hatab2022,Hatab2023}, which use line standards of different lengths, thru connection, and symmetric unknown reflect standard. The thru standard in TRL is fully defined. However, there is an implementation that eliminates the requirement of the thru standard for any transmissive device with an additional reflect standard \cite{Hatab2023b}. While multiline TRL is a very accurate calibration method, especially at millimeter-wave frequencies, it cannot be applied at lower frequencies, as it results in using an extremely long line standard. A common replacement for the multiline TRL method for on-wafer application is the line-reflect-match (LRM), thru-match-reflect-reflect (TMRR), and line-reflect-reflect-match (LRRM) methods \cite{Eul1988,Zhao2017,Rumiantsev2018,Hayden2006}. These methods use unknown symmetric reflect standards and one known match standard. However, these methods suffer from some impracticality, especially in defining the line standard and shifting the reference plane, as opposed to the TRL method. These methods can also be extended to account for crosstalk \cite{Williams2014,Dahlberg2014,Wang2023}. Additionally, due to the requirement of defining the thru/line standard, such methods can be challenging to use in on-wafer measurement scenarios where the probes are orthogonal or at an angle \cite{Basu1997}.

In this paper, we propose a new approach to self-calibration of VNAs using multiple symmetric one-port loads, a two-port reciprocal device, and a matched load. The multi-load one-port standards are two-port symmetric loads, and at least three unique loads must be used. The values of the loads themselves are not specified. For example, a short, an open, and any finite impedance load would be suitable. The reciprocal device can be any transmissive device. In fact, if we only care about the one-port error boxes of the VNA, then the two-port device can be any transmissive device, even if it is non-reciprocal. Lastly, the matched load is fully defined but can be asymmetric. The match standard can be implemented as part of the symmetric one-port loads to reduce the number of standards. We refer to this calibration method as the symmetric-reciprocal-match (SRM) method. All standards are generally partially defined, except for the match standard. We demonstrate the method using synthetic data of coplanar waveguide (CPW) structures, as well as measurements with commercial SOLR coaxial standards.

A significant benefit of the proposed approach is that all the standards are partially defined, except for the match standard. This is in contrast to LRRM/LRM/TMRR approaches, which necessitate fully defined thru/line standards. As a result, such techniques can be challenging in the case of on-wafer setups where the probes are positioned at an orthogonal angle. Equivalently, the SOLR calibration addresses the problem of the thru/line connection by using any two-port reciprocal device instead but necessitates the specification of the remaining standards. In brief, our SRM technique combines the benefits of LRRM/LRM/TMRR techniques in utilizing undefined symmetric standards, as well as the SOLR technique in utilizing a two-port reciprocal device. This revised definition of the standards enables accurate calibration by limiting the definition to solely the match standard.

The remainder of this article is structured as follows. In Section~\ref{sec:2}, we discuss our SRM method when using a thru standard instead of any reciprocal device, highlighting the method's fundamentals. Afterward, in Section~\ref{sec:3}, we extend the mathematics of the calibration to consider any transmissive reciprocal device. Section~\ref{sec:4} introduces a special case of the SRM method when considering a fixed distance between measuring ports, which is often the case in on-wafer applications. Lastly, in Section~\ref{sec:5}, we provide experimental measurements using commercial METAS traceable coaxial $2.92\,\mathrm{mm}$ calibration and verification standards, as well as numerical Monte Carlo analysis using synthetic data. Finally, we conclude in Section~\ref{sec:6}.

% EOF
\section{The Simple Case Using a Thru Standard}
\label{sec:2}

In the general case of SRM calibration, no thru standard is required. Any transmissive reciprocal device would suffice. If only the one-port error boxes are desired, any transmissive device would be acceptable. However, the derivation of the generalized SRM calibration is based on creating an artificial thru standard via mathematical reformulation and additional one-port measurements. The handling of the artificial thru standard is explained in more detail in Section~\ref{sec:3}. In this section, we assume a fully defined thru standard to derive the calibration workflow and extend it to the general case in Section~\ref{sec:3}.

\begin{figure}[th!]
	\centering
	\includegraphics[width=0.95\linewidth]{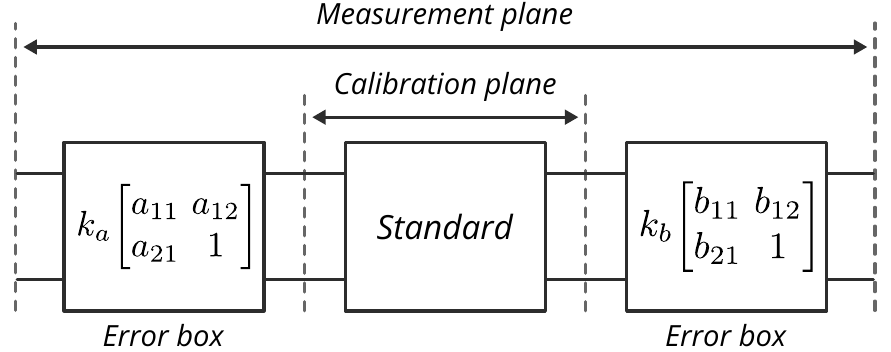}
	\caption{Two-port VNA error box model. Matrices are given as T-parameters.}
	\label{fig:2.1}
\end{figure}

To start the derivation, we use the error box model of a two-port VNA, as illustrated in Fig.~\ref{fig:2.1}. This model is expressed in T-parameters as follows:
\begin{equation}
	\bs{M}_\mathrm{stand} = \underbrace{k_ak_b}_{k}\underbrace{\left[\begin{matrix}a_{11} & a_{12}\\[5pt]\
			a_{21} & 1\end{matrix}\right]}_{\bs{A}}\bs{T}_\mathrm{stand} \underbrace{\left[\begin{matrix}b_{11} & b_{12}\\[5pt]
			b_{21} & 1\end{matrix}\right]}_{\bs{B}}, 
	\label{eq:2.1}
\end{equation}
where $\bs{M}_\mathrm{stand}$ and $\bs{T}_\mathrm{stand}$ represent the measured and actual T-parameters of the standard, respectively. The matrices $\bs{A}$ and $\bs{B}$ are the one-port error boxes containing the first six error terms, and $k$ is the seventh error term that describes the transmission error between the ports.

For a thru standard, the measured T-parameters are provided as follows:
\begin{equation}
	\bs{M}_\mathrm{thru} = k\bs{A}\bs{B}.
	\label{eq:2.2}
\end{equation}

\begin{figure}[th!]
	\centering
	\includegraphics[width=0.95\linewidth]{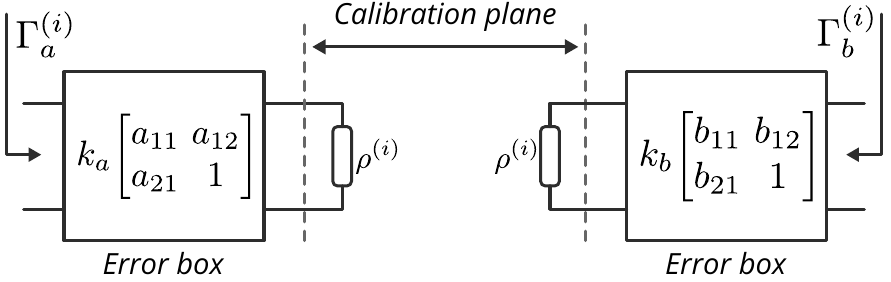}
	\caption{Two-port VNA error box model that illustrates the measurement of one-port standards. All matrices are provided as T-parameters. The index $i$ indicates the measured standard, where $i=1,2,\ldots, M$, with $M \geq 3$.}
	\label{fig:2.2}
\end{figure}

In the next step, we will focus on measuring one-port standards. For the SRM method, we require at least three symmetric two-port standards made from one-port devices, and at least three of them should exhibit unique electrical responses. Examples of such standards include short, open, and impedance. It is not necessary to know the exact response of the standards themselves. Fig.~\ref{fig:2.2} provides an illustration of the error box for one-port measurements.

The measured input reflection coefficient seen from each port is given as follows:
\begin{equation}
	\Gamma_a^{(i)} = \frac{a_{11}\rho^{(i)}+a_{12}}{a_{21}\rho^{(i)}+1}; \quad \Gamma_b^{(i)} = \frac{b_{11}\rho^{(i)}-b_{21}}{1-b_{12}\rho^{(i)}},
	\label{eq:2.3}
\end{equation}
where $\Gamma_a^{(i)}$ and $\Gamma_b^{(i)}$ are the $i$th measured reflection coefficients from the left and right ports, respectively. The actual response of the standard, which is assumed to be unknown, is denoted by $\rho^{(i)}$.

The expression for the input reflection coefficient, as given in \eqref{eq:2.3}, is in the form of a Möbius transformation (also known as a bilinear transformation) \cite{Needham2023}. One important property of the Möbius transformation is that it can be described by an equivalent $2\times2$ matrix notation. For instance, \eqref{eq:2.4} provides a general Möbius transformation with coefficients $a,b,c,d\in\mathbb{C}$, along with its corresponding $2\times 2$ matrix representation.
\begin{equation}
	f(z) = \frac{az+b}{cz+d} \quad \longleftrightarrow \quad[f]=\begin{bmatrix}
		a & b\\
		c & d
	\end{bmatrix}
	\label{eq:2.4}
\end{equation}

In \eqref{eq:2.4}, we use brackets $[\cdot]$ to describe matrices associated with a Möbius transformation. The transformation coefficients are only unique up to a complex scalar multiple. This property of the Möbius transform can be easily shown by multiplying the numerator and denominator with a non-zero complex scalar. In terms of matrix notation, scaling the matrix with a complex scalar still represents the same Möbius transformation. Therefore,
\begin{equation}
	\quad[f] \equiv \kappa[f], \quad \kappa\neq 0
	\label{eq:2.5}
\end{equation}

The matrix representation of the Möbius transformation possesses an elegant property in its ability to describe composite Möbius transformations. In essence, when we compose one Möbius transformation with another, we obtain a new Möbius transformation with updated coefficients. This property can be expressed in matrix notation by computing the matrix product of the individual Möbius transformations. To illustrate this concept, we provide an example of the composition of two Möbius transformations $f_1(z)$ and $f_2(z)$, which are defined as follows:
\begin{subequations}
\begin{align}
		f_1(z) &= \frac{a_1z+b_1}{c_1z+d_1} \quad \longleftrightarrow \quad[f_1]=\begin{bmatrix}
		a_1 & b_1\\
		c_1 & d_1
	\end{bmatrix}\\[5pt]
		f_2(z) &= \frac{a_2z+b_2}{c_2z+d_2} \quad \longleftrightarrow \quad [f_2]=\begin{bmatrix}
			a_2 & b_2\\
			c_2 & d_2
		\end{bmatrix}
\end{align}
	\label{eq:2.6}
\end{subequations}

The composite transformation is given as follows:
\begin{equation}
	\begin{aligned}
		g(z) = (f_1\circ f_2)(z) &= \frac{a_1f_2(z)+b_1}{c_1f_2(z)+d_1}\\[5pt]
		&= \frac{(a_1a_2 + b_1c_2)z + a_1b_2 + b_1d_2}{(a_2c_1 + c_2d_1)z + b_2c_1 + d_1d_2}
	\end{aligned}
	\label{eq:2.7}
\end{equation}

Therefore, the corresponding matrix equivalent of the composite Möbius transformation $g(z)$ is given as follows:
\begin{equation}
	[g]=\begin{bmatrix}a_{1} a_{2} + b_{1} c_{2} & a_{1} b_{2} + b_{1} d_{2}\\[5pt]
	a_{2} c_{1} + c_{2} d_{1} & b_{2} c_{1} + d_{1} d_{2}\end{bmatrix} = [f_1][f_2]
	\label{eq:2.8}
\end{equation}
which is the same as multiplying the matrices $[f_1]$ and $[f_2]$.

Using matrix notation for the Möbius transformation, we can describe the input reflection coefficient measured from the left port as follows:
\begin{equation}
	\Gamma_a^{(i)} = \frac{a_{11}\rho^{(i)}+a_{12}}{a_{21}\rho^{(i)}+1} \longleftrightarrow
	[\Gamma_a^{(i)}]  = \underbrace{\begin{bmatrix}
		a_{11} & a_{12}\\[5pt]
		a_{21} & 1
	\end{bmatrix}}_{\bs{A}}
	\label{eq:2.9}
\end{equation}

To address the error box on the right side, we perform a similar process as before, but instead of using the measured reflection coefficient, we reformulate in terms of the reflection coefficient $\rho^{(i)}$ as a function of the measured reflection coefficient $\Gamma_b^{(i)}$, which is given as follows:
\begin{equation}
	\rho^{(i)} = \frac{\Gamma_b^{(i)}+b_{21}}{b_{12}\Gamma_b^{(i)}+b_{11}} \longleftrightarrow
	[\rho^{(i)}]  = \underbrace{\begin{bmatrix}
			1 & b_{21}\\[5pt]
			b_{12} & b_{11}
	\end{bmatrix}}_{\bs{P}\bs{B}\bs{P}}
	\label{eq:2.10}
\end{equation}
where $\bs{P}$ is a $2\times 2$ permutation matrix defined as 
\begin{equation}
	\bs{P} = \bs{P}^T = \bs{P}^{-1} = \begin{bmatrix}
		0 & 1\\
		1 & 0
	\end{bmatrix}.
	\label{eq:2.11}
\end{equation}

By composing \eqref{eq:2.10} with \eqref{eq:2.9}, we obtain a new Möbius transformation that describes the input reflection coefficient from the left port using measurements of the right port. This relationship can be written as follows:
\begin{equation}
	\Gamma_a^{(i)} = \frac{h_{11}\Gamma_b^{(i)}+h_{12}}{h_{21}\Gamma_b^{(i)}+h_{22}} \longleftrightarrow [\Gamma_a^{(i)}]=\bs{H}=\begin{bmatrix}
		h_{11} & h_{12}\\
		h_{21} & h_{22}
	\end{bmatrix}
	\label{eq:2.12}
\end{equation}

Here, we use the variable $\bs{H}$ to describe the Möbius transformation in \eqref{eq:2.12} and differentiate it from the Möbius transformation in \eqref{eq:2.9} to avoid confusion. It is important to note that both transformations are different, as they have distinct input parameters.

Due to the composite property of Möbius transformations, the coefficients of the transformation can be expressed as follows:
\begin{equation}
	\bs{H} = \nu\bs{A}\bs{P}\bs{B}\bs{P}, \qquad \forall\,\nu \neq 0.
	\label{eq:2.13}
\end{equation}

It is important to note that the constant $\nu$ is included because the coefficients of a Möbius transformation can only be defined up to a non-zero complex-valued scalar constant.

By solving for the coefficients $h_{ij}$, we can determine \eqref{eq:2.13}. This equation is later used for establishing the calibration procedure by combining it with the thru standard. Since the coefficients $h_{ij}$ are defined by the Möbius transformation in \eqref{eq:2.12}, which is based on the measurements of the symmetric one-port standards, we can rewrite the Möbius transformation as a linear system of equations in terms of its coefficients. Assuming that $M\geq3$ one-port standards were measured, the coefficients $h_{ij}$ can be described as follows:
\begin{equation}
	\underbrace{\left[\begin{matrix}
			-\Gamma_b^{(1)} & -1 & \Gamma_b^{(1)}\Gamma_a^{(1)} & \Gamma_a^{(1)} \\
			-\Gamma_b^{(2)} & -1 & \Gamma_b^{(2)}\Gamma_a^{(2)} & \Gamma_a^{(2)}\\
			\vdots & \vdots & \vdots & \vdots\\
			-\Gamma_b^{(M)} & -1 & \Gamma_b^{(M)}\Gamma_a^{(M)} & \Gamma_a^{(M)}
		\end{matrix}\right]}_{\bs{G}}
	\underbrace{\left[\begin{matrix}
			h_{11}\\
			h_{12}\\
			h_{21}\\
			h_{22}
		\end{matrix}\right]}_{\bs{h}} = \bs{0}
	\label{eq:2.14}
\end{equation}

The solution for the vector $\bs{h}$ is found in the nullspace of $\bs{G}$, as the system matrix $\bs{G}$ contains at least one nullspace due to the equality to zero in \eqref{eq:2.14}. We may have more than one nullspace, but only if $\mathrm{rank}(\bs{G}) < 3$, which can only happen if we do not use at least three unique one-port standards.

While the nullspace $\bs{G}$ satisfies the solution of \eqref{eq:2.14}, we can optimally estimate the nullspace of $\bs{G}$ in the presence of disturbance by computing its singular value decomposition (SVD) and using the right singular vector that corresponds to the smallest singular value \cite{Strang1993}. As $\bs{G}$ is of dimension 4 (i.e., number of columns), it has four singular values and vectors. We decompose the matrix $\bs{G}$ using SVD as follows:
\begin{equation}
	\bs{G} = \sum_{i=1}^{4} s_i\bs{u}_i\bs{v}_i^{H}
	\label{eq:2.15}
\end{equation}
where $s_i$ is the $i$th singular value, while $\bs{u}_i$ and $\bs{v}_i$ are the $i$th left and right singular vectors, respectively. The conventional ordering of the singular values is in decreasing order. Therefore, the smallest singular value is $s_4$. Hence, the solution for $\bs{h}$ is given by the fourth right singular vector as follows:
\begin{equation}
	\bs{h} = \bs{v}_4
	\label{eq:2.16}
\end{equation}

Now that we have solved for $\bs{h}$, and hence $\bs{H}$ in \eqref{eq:2.13}, we can combine the measurements of the thru standards with the results of $\bs{H}$ to form an eigenvalue problem regarding the error box coefficients. The combined result for the left error box is defined as follows:
\begin{equation}
	\bs{M}_\mathrm{thru}\bs{P}\bs{H}^{-1} = \frac{k}{\nu} \bs{A}\bs{P}\bs{A}^{-1}
	\label{eq:2.17}
\end{equation}

Although \eqref{eq:2.17} is not strictly in the canonical form for an eigenvalue decomposition, as the middle matrix is not diagonal, it can still be decomposed because the middle matrix is a constant permutation matrix. If we apply the eigendecomposition to \eqref{eq:2.17}, we obtain the following decomposition:
\begin{equation}
	\bs{M}_\mathrm{thru}\bs{P}\bs{H}^{-1} = \frac{k}{\nu} \bs{A}\bs{P}\bs{A}^{-1} = \bs{W}_a\bs{\Lambda}\bs{W}_a^{-1},
	\label{eq:2.18}
\end{equation}
where the matrix $\bs{W}_a$ corresponds to the eigenvectors, and the matrix $\bs{\Lambda}$ corresponds to the eigenvalues. Both are calculated as follows:
\begin{subequations}
\begin{align}
	\bs{W}_a &= \begin{bmatrix}
				w_{11}^{(a)} & w_{12}^{(a)}\\[5pt]
		w_{21}^{(a)} & w_{22}^{(a)}
	\end{bmatrix} = \begin{bmatrix}
	\frac{a_{11}+a_{12}}{a_{21}+1} & \frac{-a_{11}+a_{12}}{-a_{21}+1}\\[5pt]
	1 & 1
	\end{bmatrix}\\[5pt]
	\bs{\Lambda} &= \begin{bmatrix}
		\lambda_1 & 0 \\[5pt]
		0 & \lambda_2
	\end{bmatrix} = \begin{bmatrix}
		\frac{k}{\nu} & 0 \\[5pt]
		0 & -\frac{k}{\nu}
	\end{bmatrix}
\end{align}
\label{eq:2.19}
\end{subequations}

Generally, the order of the eigenvectors and eigenvalues is not unique. To ensure the correct order, we need to know the value of $k/\nu$. However, this term is still unknown at this stage. After solving for the error terms using both possible solutions, the sorting is done through trial and error. For instance, once the error terms have been solved, we could use one of the one-port standards as a metric to determine the correct order.

We can solve the eigenvalue problem for matrix $\bs{B}$ by reversing the multiplication order of the matrices in \eqref{eq:2.17}. This gives us the following equation:
\begin{equation}
	\left( \bs{P}\bs{H}^{-1}\bs{M}_\mathrm{thru}\right)^T = \frac{k}{\nu} \bs{B}^T\bs{P}\bs{B}^{-T} = \bs{W}_b\bs{\Lambda}\bs{W}_b^{-1}
	\label{eq:2.20}
\end{equation}

Using the transpose operation is optional, but it allows us to derive the eigenvectors in a similar order as with the left error box. As a result, the eigenvectors and eigenvalues are given as follows:
\begin{subequations}
	\begin{align}
		\bs{W}_b &= \begin{bmatrix}
			w_{11}^{(b)} & w_{12}^{(b)}\\[5pt]
			w_{21}^{(b)} & w_{22}^{(b)}
		\end{bmatrix} = \begin{bmatrix}
			\frac{b_{11}+b_{21}}{b_{12}+1} & \frac{-b_{11}+b_{21}}{-b_{12}+1}\\[5pt]
			1 & 1
		\end{bmatrix}\\[5pt]
		\bs{\Lambda} &= \begin{bmatrix}
			\lambda_1 & 0 \\[5pt]
			0 & \lambda_2
		\end{bmatrix} = \begin{bmatrix}
			\frac{k}{\nu} & 0 \\[5pt]
			0 & -\frac{k}{\nu}
		\end{bmatrix}
	\end{align}
	\label{eq:2.21}
\end{subequations}

Finally, we need an additional equation for each port to calculate the error terms from each error box. This equation comes from the match standard, which defines the reference impedance of the calibration. In general, the match standard does not have to be the same at each port. However, since we are most likely to use an impedance standard as part of the symmetric one-port devices, it makes sense to reuse the match standards. For each port, the reflection coefficient of a match standard is given as follows:
\begin{equation}
	\rho_a^{(m)} = \frac{Z_a^{(m)}-Z_a^{(ref)}}{Z_a^{(m)}+Z_a^{(ref)}}; \quad \rho_b^{(m)} = \frac{Z_b^{(m)}-Z_b^{(ref)}}{Z_b^{(m)}+Z_b^{(ref)}}
	\label{eq:2.22}
\end{equation}
where $Z_a^{(m)}$ and $Z_b^{(m)}$ represent the complex impedance definition of the match standard from each port. The user sets the values of $Z_a^{(ref)}$ and $Z_b^{(ref)}$ to specify the reference impedance, for example, $50\,\Omega$.

By utilizing knowledge of the match standard and the equation that describes the input reflection coefficient, as given in \eqref{eq:2.3}, we can combine this result with the eigenvectors to form a linear system of equations for each port. The following is for the left port:
\begin{equation}
	\left[\begin{matrix}
			-1 & -1 & w_{11}^{(a)} & w_{11}^{(a)} \\
			1 & -1 & -w_{12}^{(a)} & w_{12}^{(a)} \\
			-\rho_a^{(m)} & -1 & \Gamma_a^{(m)}\rho_a^{(m)} & \Gamma_a^{(m)}
		\end{matrix}\right]
	\left[\begin{matrix}
			a_{11}\\
			a_{12}\\
			a_{21}\\
			1
		\end{matrix}\right] = \bs{0}
	\label{eq:2.23}
\end{equation}

The system of equations for the right port can be obtained in a similar way, resulting in the following system of equations:
\begin{equation}
	\left[\begin{matrix}
		-1 & -1 & w_{11}^{(b)} & w_{11}^{(b)} \\
		1 & -1 & -w_{12}^{(b)} & w_{12}^{(b)} \\
		-\rho_b^{(m)} & 1 & -\Gamma_b^{(m)}\rho_b^{(m)} & \Gamma_b^{(m)}
	\end{matrix}\right]
	\left[\begin{matrix}
		b_{11}\\
		b_{21}\\
		b_{12}\\
		1
	\end{matrix}\right] = \bs{0}
	\label{eq:2.24}
\end{equation}

The error terms are solved by finding the nullspace of the system matrix. However, since the nullspace is only unique up to a scalar factor, we normalize it by the last element to make it equal to 1. The system matrix can be extended by an arbitrary number of defined impedance standards to improve the solution. It is important to note that we obtain two systems of equations for each port since the order of the eigenvectors is unknown. As a result, we solve for both possible orderings and choose the answer that results in a calibrated measurement closest to a known estimate, like the usage of a reflect standard.

An interesting observation to note is the structure of \eqref{eq:2.23} and \eqref{eq:2.24}, where the first two rows in the system matrix obtained from the eigenvectors resemble measurements of ideal short and open standards. In general, the expression of \eqref{eq:2.23} and \eqref{eq:2.24} are identical to that of a one-port SOL calibration when assuming ideal short and open standards. Thus, we were able to replicate measurements of ideal open and short standards by using symmetric undefined one-port devices and a thru standard.

The final error term that needs to be solved is the transmission error term $k$. Since we are working with a thru standard, we can directly extract $k$ by multiplying the inverse of the one-port error boxes by the measurements of the thru standard. In Section~\ref{sec:3}, we introduce a different approach for computing $k$ using any transmissive reciprocal standard, as done in SOLR calibration \cite{Ferrero1992}.

% EOF
\section{Generalization without a Thru Standard}
\label{sec:3}

In the previous section, we explained how to calculate the error terms using at least three symmetric one-port standards, a thru standard, and a match standard. The thru standard can cause difficulties, as it is not always possible to physically achieve such a standard. 

The equations derived in the previous section can be used without changes if we obtain an equation similar to that of a thru standard, as given in \eqref{eq:2.2}. Therefore, this section aims to derive what we will refer to as a virtual thru standard by using additional one-port standards.

\begin{figure}[th!]
	\centering
	\includegraphics[width=0.95\linewidth]{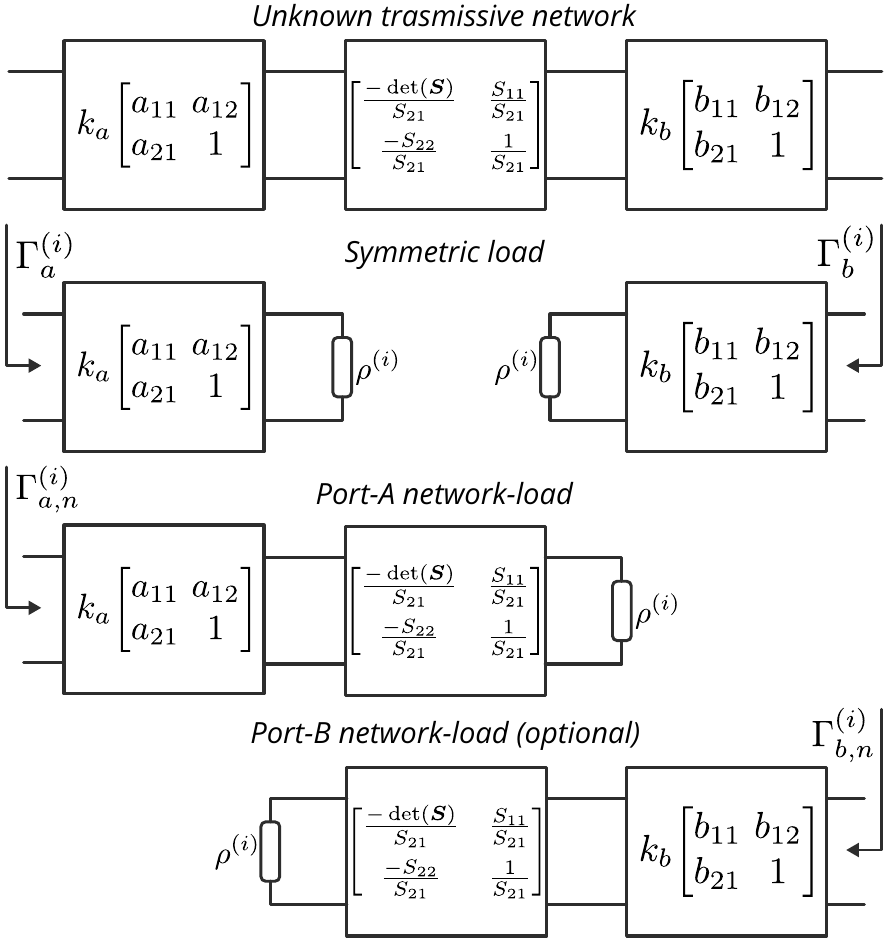}
	\caption{Two-port VNA error box model illustrating the standards used to create a virtual thru standard. All matrices are provided as T-parameters. The index $i$ indicates the measured standard, where $i=1,2,\ldots, M$, with $M \geq 3$.}
	\label{fig:3.1}
\end{figure}

The necessary standards, excluding the match standard, for the generalized SRM calibration are shown in Fig.~\ref{fig:3.1}. The network standard is an unknown transmissive two-port standard. This standard does not need to be reciprocal for deriving only one-port error terms. The additional network-load standard uses the same two-port network standard and the same one-port symmetric standards. As mentioned in the previous section, we require at least $M\geq3$ one-port symmetric standards. Hence, we also need a corresponding network-load standard for every symmetric one-port load standard. Generally, we only need the network-load standard from one port, which could be from either ports.

Based on the network standard, the following measurement is available:
\begin{equation}
	\bs{M}_\mathrm{net} = k\bs{A}\underbrace{\begin{bmatrix}
		\frac{-\det\left(\bs{S}\right)}{S_{21}} & \frac{S_{11}}{S_{21}}\\[5pt]
		\frac{-S_{22}}{S_{21}} & \frac{1}{S_{21}}
	\end{bmatrix}}_{\bs{N}}\bs{B}
	\label{eq:3.1}
\end{equation}
where $\det\left(\bs{S}\right) = S_{11}S_{22}-S_{12}S_{21}$.

A similar expression to the matrix $\bs{H}$ in \eqref{eq:2.13} can be obtained using the network-load standard from the left port and the load standards from the right ports. This results in an expression similar to \eqref{eq:2.13}, but with $\bs{A}$ replaced by $\bs{A}\bs{N}$ and with an adjustment to the scaling factor. The scaling factor is unknown and does not need to be equal to the constant in \eqref{eq:2.13}. We can also achieve the same result by considering the network-load standards from the right port and symmetric load standards from the left port. As a result, combining the network-load standards with the symmetric load standards, we obtain the following result for each port depending on where the network-load was implemented:
\begin{subequations}
	\begin{align}
		\bs{F}_a &= \eta\bs{A}\bs{N}\bs{P}\bs{B}\bs{P}, \qquad \forall\,\eta \neq 0,\label{eq:3.3a}\\
		\bs{F}_b &= \zeta\bs{A}\bs{P}\bs{N}\bs{B}\bs{P}, \qquad \forall\,\zeta \neq 0\label{eq:3.3b}
	\end{align}
	\label{eq:3.3}
\end{subequations}

Using the results of $\bs{M}_\mathrm{net}$, $\bs{H}$, and $\bs{F}$ from \eqref{eq:3.1}, \eqref{eq:2.13}, and \eqref{eq:3.3}, respectively, we can create a virtual thru standard by combining them in the following manner:
\begin{subequations}
	\begin{align}
		\bs{M}_\mathrm{thru} &= \bs{H}\bs{F}_a^{-1}\bs{M}_\mathrm{net} =\frac{\nu}{\eta}k\bs{A}\bs{B}\\
		\bs{M}_\mathrm{thru} &= \bs{M}_\mathrm{net}\bs{P}\bs{F}_b^{-1}\bs{H}\bs{P} =\frac{\nu}{\zeta}k\bs{A}\bs{B}
	\end{align}
	\label{eq:3.4}
\end{subequations}

Therefore, we can obtain a thru measurement without measuring a thru standard using the results of \eqref{eq:3.4}. We simply use the results from the previous section and substitute \eqref{eq:3.4} in place of the thru measurements. The only difference we obtain are the eigenvalues, which result in $\pm k/\eta$ or $\pm k/\zeta$. However, this change does not affect anything, as $\nu$, $\eta$, and $\zeta$ are the result of the normalization choice of the Möbius transformation and are assumed regardless unknown.

To complete the two-port calibration, we must solve for the transmission error term $k$. We can use the same method as in SOLR calibration \cite{Ferrero1992} by calculating $k$ through the determinate of the one-port corrected measurement of the network standard, given that it is reciprocal (i.e., $S_{21}=S_{12}$). Assuming the network standard is indeed reciprocal, we can solve for $k$ by first applying the one-port error boxes to the measurement of the network standard as follows:
\begin{equation}
	\bs{A}^{-1}\bs{M}_\mathrm{net}\bs{B}^{-1} = k\bs{N}
	\label{eq:3.5}
\end{equation}

Afterward, by taking the determinant from both sides, we obtain the following:
\begin{equation}
	\det\left(\bs{A}^{-1}\bs{M}_\mathrm{net}\bs{B}^{-1}\right) = k^2\underbrace{\det\left(\bs{N}\right)}_{=1}
	\label{eq:3.6}
\end{equation}

Hence, $k$ is solved as follows:
\begin{equation}
	k = \pm\sqrt{\det\left(\bs{A}^{-1}\bs{M}_\mathrm{recip}\bs{B}^{-1}\right)}
	\label{eq:3.7}
\end{equation}
where the selection of the appropriate sign is determined by comparing it to a known estimate of the network.

% EOF
\section{Special Layout for On-wafer Application}
\label{sec:4}

The presented SRM calibration method applies to any measurement setup where the standards can be implemented. However, a particular case for on-wafer calibration arises when considering that the distance between the probes must remain constant. Semi-automatic probe station users often request this requirement, where only the chuck platform is motorized. For these measurement setups, the standards must be implemented with a constant distance between the probes to perform the calibration automatically.

Considering the standards depicted in Fig.~\ref{fig:3.1}, we can see that the right probe would need to be moved to the right to measure the network-load standard. The network standard already dictates the distance between the probes, and cascading another standard would naturally increase the spacing, requiring probe movement.

\begin{figure}[th!]
	\centering
	\includegraphics[width=1\linewidth]{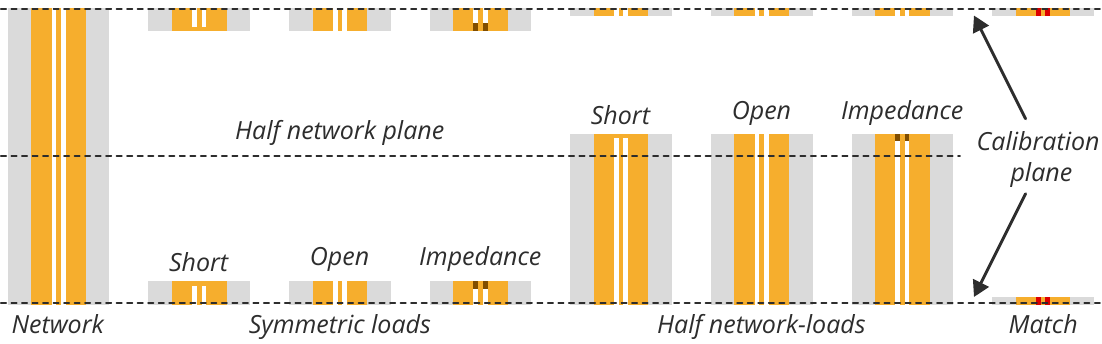}
	\caption{Illustration of CPW structures implementing the proposed half-network approach of SRM calibration. The match standard is optional if the symmetric impedance standard is reused as the match standard.}
	\label{fig:4.1}
\end{figure}

In planar circuit calibration, as in on-wafer measurement setups, we can advantageously apply the property of the network standard to represent any symmetric transmissive network. Hence, we can split the network into two cascaded flipped asymmetric networks. With this notation, we can use half of the network to define the network-load standard.  An illustration of coplanar waveguide (CPW) standards is depicted in Fig.~\ref{fig:4.1}.

For any symmetric network (i.e., $S_{ij}=S_{ji}$), we can divide its T-parameters into two cascaded networks that are identical and flipped \cite{Marks1992}. This network can be expressed as follows:
\begin{equation}
	\bs{N} = \underbrace{\bs{R}}_{\text{left half}}\underbrace{\bs{P}\bs{R}^{-1}\bs{P}}_{\text{right half}}
	\label{eq:4.1}
\end{equation}
where $\bs{P}$ represents the permutation matrix, as defined in \eqref{eq:2.11}, and $\bs{R}$ is the half-asymmetric part of the network standard. 

By substituting \eqref{eq:4.1} into \eqref{eq:3.1}, and the right and left half networks into \eqref{eq:3.3a} and \eqref{eq:3.3b}, respectively, we obtain the following expressions:
\begin{subequations}
	\begin{align}
		\bs{M}_\mathrm{net} &= k\bs{A}\bs{R}\bs{P}\bs{R}^{-1}\bs{P}\bs{B}\\
		\bs{F}_a &= \eta\bs{A}\bs{R}\bs{P}\bs{B}\bs{P}, \qquad \forall\,\eta \neq 0\\
		\bs{F}_b &= \zeta\bs{A}\bs{R}^{-1}\bs{P}\bs{B}\bs{P}, \qquad \forall\,\zeta \neq 0.
	\end{align}
	\label{eq:4.2}
\end{subequations}

Therefore, by combining the results of the above expressions with $\bs{H}$ from \eqref{eq:2.13}, we create a virtual thru standard as follows:
\begin{subequations}
	\begin{align}
		\bs{M}_\mathrm{thru} &= \bs{H}\bs{F}_a^{-1}\bs{M}_\mathrm{net}\bs{P}\bs{H}^{-1}\bs{F}_a\bs{P} = k\bs{A}\bs{B},\\
		\bs{M}_\mathrm{thru} &= \bs{F}_b\bs{H}^{-1}\bs{M}_\mathrm{net}\bs{P}\bs{F}_b^{-1}\bs{H}\bs{P} = k\bs{A}\bs{B}.
	\end{align}
	\label{eq:4.3}
\end{subequations}

With the virtual thru standard being established, the remaining calibration process follows the same procedure discussed in the previous section.

\begin{figure}[th!]
	\centering
	\includegraphics[width=1\linewidth]{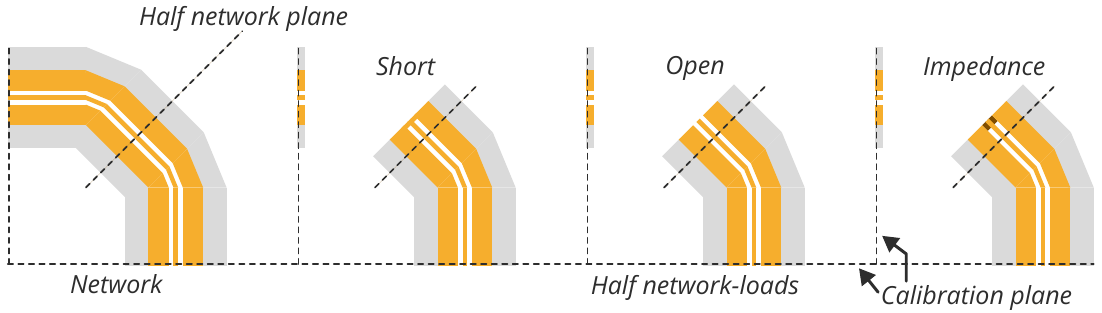}
	\caption{Illustration of CPW structures implementing the half network-load standards in an orthogonal orientation. The symmetric one-port standards are not shown, as they do not pose any mechanical challenge in orthogonal orientation.}
	\label{fig:4.2}
\end{figure}

One elegant application using half-network standards is the use of angled calibration. This method involves positioning the probes at an angle rather than facing each other. Traditional calibration methods such as TRL, LRM, and LRRM do not allow this type of calibration, whereas SOLR is often used for such scenarios \cite{Basu1997}. Fig.~\ref{fig:4.2} illustrates a potential implementation of the network and half-network standards at a $90^\circ$ angle.

\section{Experiments}
\label{sec:5}

Two experiments are described in this section. In the first experiment, measurements were performed using METAS traceable coaxial standards and the SRM method was compared with SOLR calibration. The second experiment demonstrates the application of the SRM method for on-wafer calibration. Since the SRM method presented here is new, there are no commercially available impedance substrate standard (ISS) kits that contain all the necessary standards, especially the network-load standards. Therefore, we decided to perform a Monte Carlo (MC) analysis using synthetic CPW data based on an actual on-wafer setup. The SRM standards were generated using a validated CPW model to analyze the impact of various uncertainties on the SRM calibration.

\subsection{Coaxial Measurements}

The measurement involves the comparison of the proposed SRM method with a SOLR calibration using a commercial METAS traceable calibration kit with a $2.92\,\mathrm{mm}$ interface. The calibration results are compared using verification standards with defined uncertainty bounds that is also traceable to METAS. The VNA used for the measurement is the ZVA from Rohde \& Schwarz (R\&S), and the calibration kit used is the ZN-Z229 $2.92\,\mathrm{mm}$ kit from R\&S. The standards used from the kit include short, open, and match standards with female interfaces, as well as two adapters with female-female and female-male interfaces of equal length. An SOLR calibration was conducted using the short, open, match standards, and female-female adapter, while assuming that the adapter standard is unknown during the SOLR calibration process.

\begin{figure}[th!]
	\centering
	\subfloat[]{\includegraphics[width=.32\linewidth]{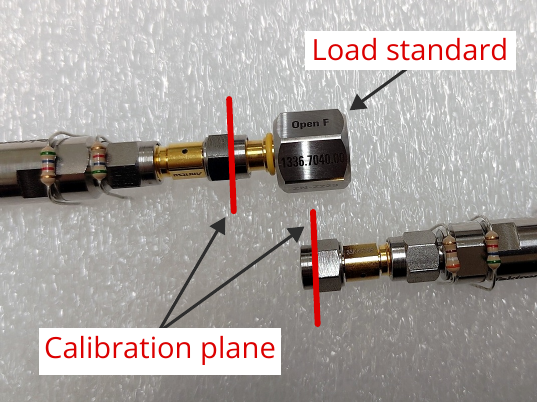}}~
	\subfloat[]{\includegraphics[width=.32\linewidth]{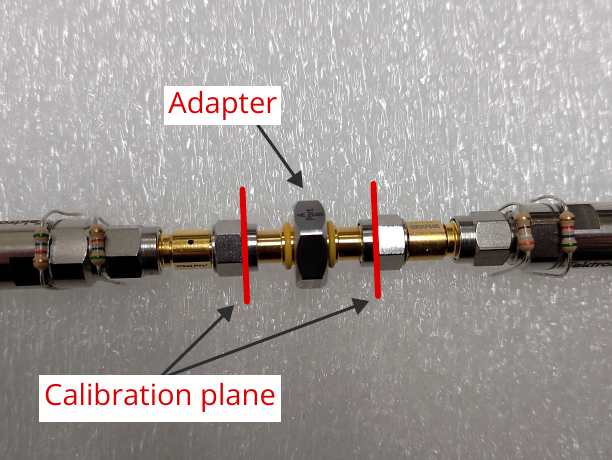}}~
	\subfloat[]{\includegraphics[width=.32\linewidth]{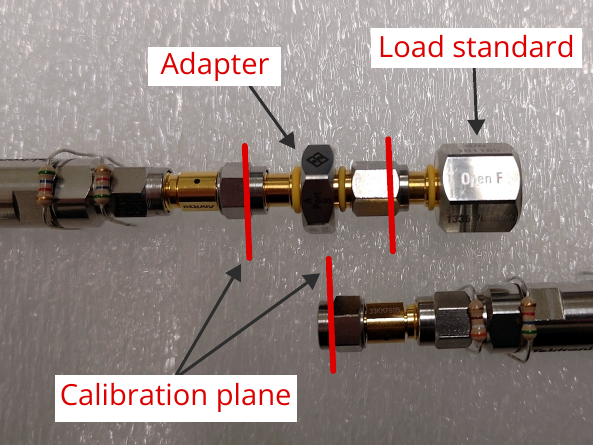}}
	\caption{Example photos of measured coaxial standards. (a) load standard, (b) adapter (network), and (c) load connected with an adapter (network-load).}
	\label{fig:5.5}
\end{figure}

For the implementation of SRM standards, the symmetrical standards are directly measured by connecting the three one-port devices at both ports. The female-female adapter is used to represent the reciprocal network. For the network-load standard, the symmetrical one-port devices are connected to the female-male adapter and measured at the left port. In all steps, the standards are assumed unknown, except for the match standard, which is only defined in the final step of the calibration via \eqref{eq:2.23} and \eqref{eq:2.24}. An example that illustrates the measurement of the standards is shown in Fig.~\ref{fig:5.5}.

The verification kit utilized for the comparison is the ZV-Z429 $2.92\,\mathrm{mm}$ kit from R\&S. The kit contains a mismatch standard and an offset short standard with female interfaces. These verification standards have been previously characterized by the manufacturer with traceability to METAS, and their S-parameters are provided with uncertainty bounds. To verify the accuracy of the calibration, we define an error metric as the magnitude of the error vector of the calibrated response to the reference response given by
\begin{equation}
	\text{Error}_{ij} \ (\mathrm{dB})  = 20\,\mathrm{log}_{10}\left| S_{ij}^\mathrm{cal} - S_{ij}^\mathrm{ref} \right|
	\label{eq:5.1}
\end{equation}
where $S_{ij}^\mathrm{cal}$ and $S_{ij}^\mathrm{ref}$ represent the calibrated and reference values, respectively. 

\begin{figure}[th!]
	\centering
	\includegraphics[width=0.95\linewidth]{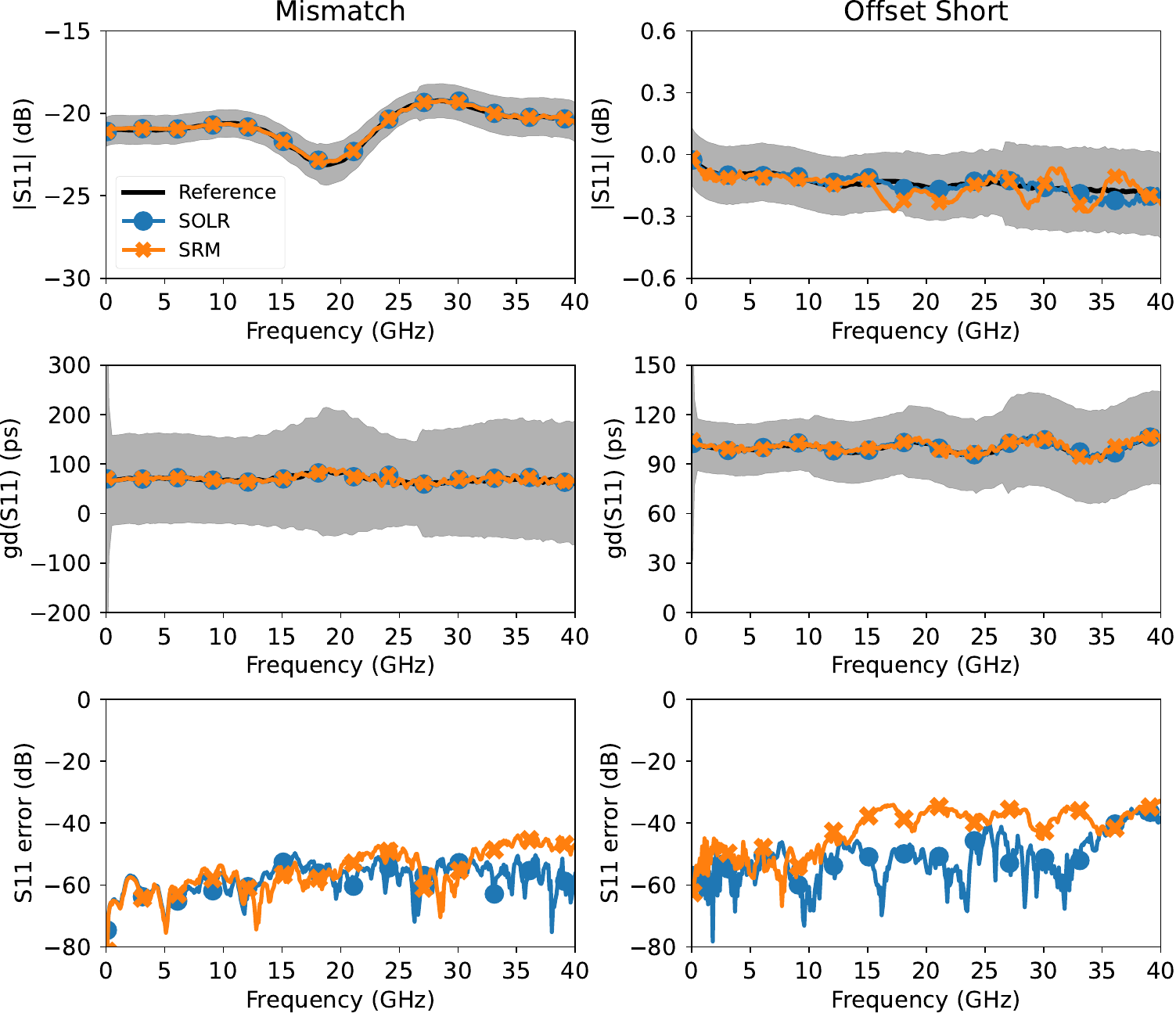}
	\caption{Comparison of calibrated mismatch and offset short verification kits using SOLR and SRM methods. The uncertainty bounds are of the reference measurement and reported as 95\% Gaussian distribution coverage.}
	\label{fig:5.6}
\end{figure}

The results from calibrating the mismatch and offset short verification kit using both SOLR and SRM calibration methods are depicted in Fig.~\ref{fig:5.6}. The plots reveal that both calibration methods produced similar outcomes, with errors relative to the reference data of the verification kit remaining below $-30\,\mathrm{dB}$. To facilitate visual comparison, we opted to plot the group delay instead of the phase. In both, the SOLR and SRM calibration, the group delay overlaps with the reference data for both mismatch and offset short. However, we observe a small discrepancy in the magnitude response of the offset short standard after $15\,\mathrm{GHz}$, where ripples can be observed. Nevertheless, this falls within the uncertainty bounds of the magnitude response of the offset short.

\begin{figure}[th!]
	\centering
	\includegraphics[width=0.85\linewidth]{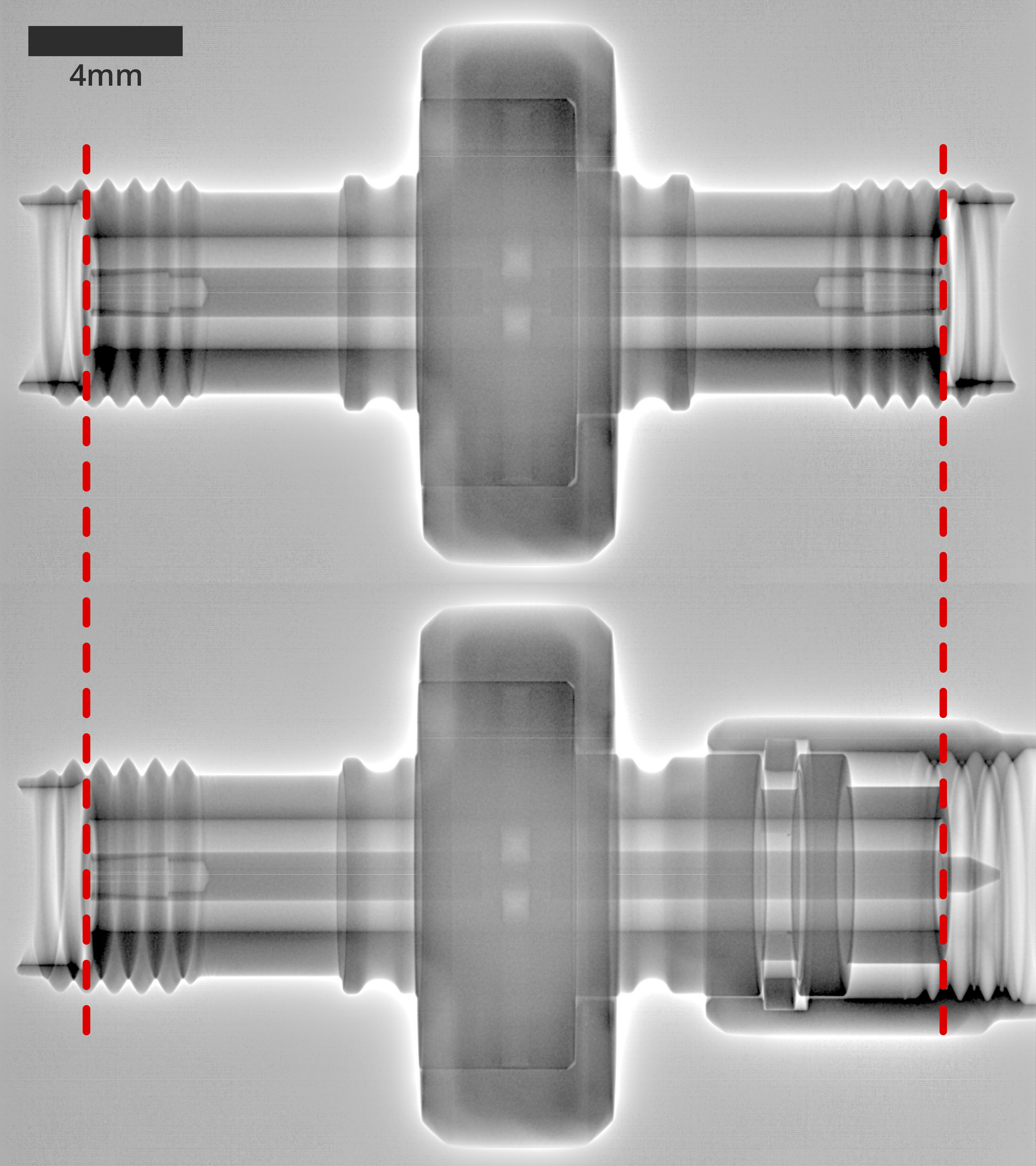}
	\caption{X-ray inspection of the female-female and female-male adapters.}
	\label{fig:5.7x}
\end{figure}

It is difficult to determine the exact cause for the ripple in the calibrated magnitude response of the offset short. This ripple is small and falls within the uncertainty bounds defined by METAS. One possible explanation for this variation could be the difference between the female-female and female-male adapters. The adapters have the same length and cross-section, as shown in the X-ray image in Fig.~\ref{fig:5.7x}. In theory, they should have the same response after pairing, as they result in a smooth continuation of the $2.92\,\mathrm{mm}$ interface \cite{IEEE2022}. However, the female interface has a slotted design, which makes this continuation not entirely smooth. Additionally, the presence of pin gaps affects different calibrations in different ways \cite{Hoffmann2007,Hoffmann2009,Wong2017,Lewandowski2019}. The pin gaps after pairing for the measured standards are summarized in Table~\ref{tab:5.1}.

\begin{table}[th!]
	\centering
	\caption{Pin gap of paired connectors. Values are reported in $\mu\mathrm{m}$. The pin depth gauge has a resolution of $2.54\,\mu\mathrm{m}$ ($0.0001\,\mathrm{in}$). The letter ``f'' stands for female (jack) and ``m'' for male (plug).}
	\label{tab:5.1}
	\begin{tabular}{@{$\quad$}cccccc@{$\quad$}}
		\toprule
		\multicolumn{1}{l}{} &
		\begin{tabular}[c]{@{}c@{}}Short\\ (f)\end{tabular} &
		\begin{tabular}[c]{@{}c@{}}Open\\ (f)\end{tabular} &
		\begin{tabular}[c]{@{}c@{}}Match\\ (f)\end{tabular} &
		\begin{tabular}[c]{@{}c@{}}Adapter\\ (ff)\end{tabular} &
		\begin{tabular}[c]{@{}c@{}}Adapter\\ (fm)\end{tabular} \\ \midrule
		\begin{tabular}[c]{@{}c@{}}Port 1\\ (m)\end{tabular}  & 31.75 & 31.75 & 31.75 & 35.56 & 54.61 \\ \midrule
		\begin{tabular}[c]{@{}c@{}}Port 2\\ (m)\end{tabular} & 31.75 & 31.75 & 31.75 & 36.83 & -     \\ \midrule
		\begin{tabular}[c]{@{}c@{}}Adapter\\ (fm)\end{tabular}    & 31.75 & 31.75 & 31.75 & -     & -     \\ \bottomrule
	\end{tabular}
\end{table}

\begin{figure}[th!]
	\centering
	\subfloat[]{\includegraphics[width=.95\linewidth]{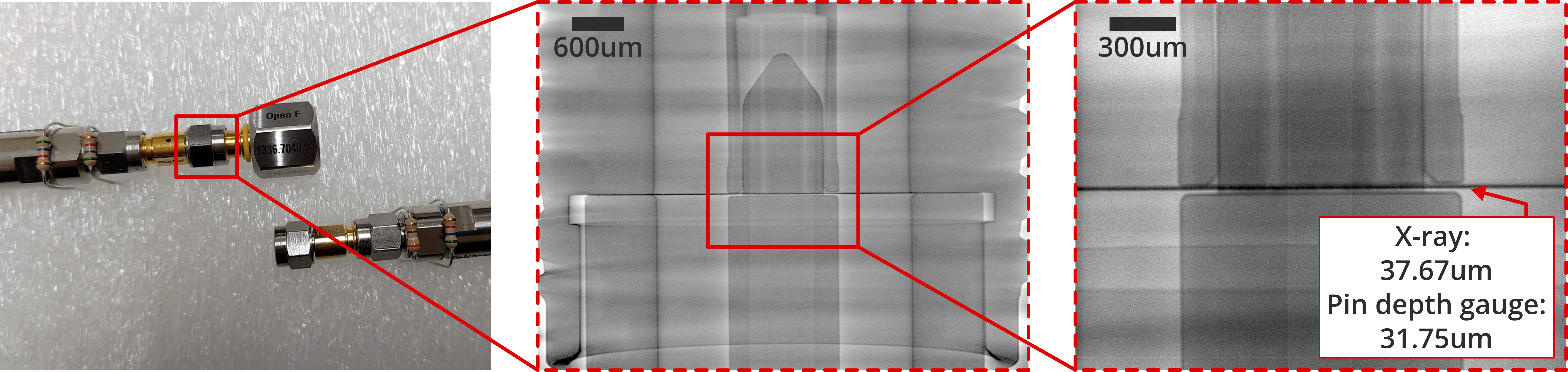}}\\
	\subfloat[]{\includegraphics[width=.95\linewidth]{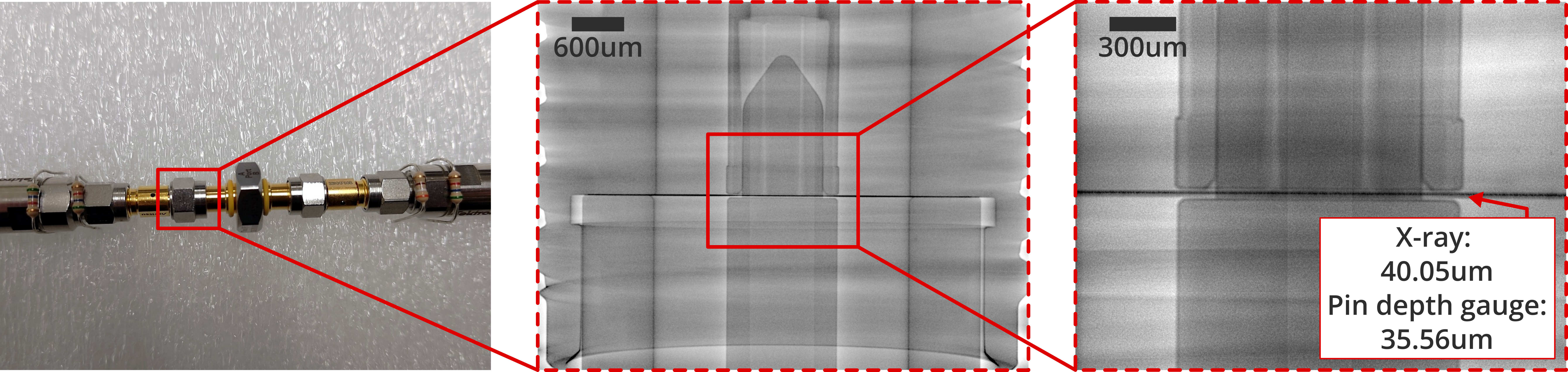}}\\
	\subfloat[]{\includegraphics[width=.95\linewidth]{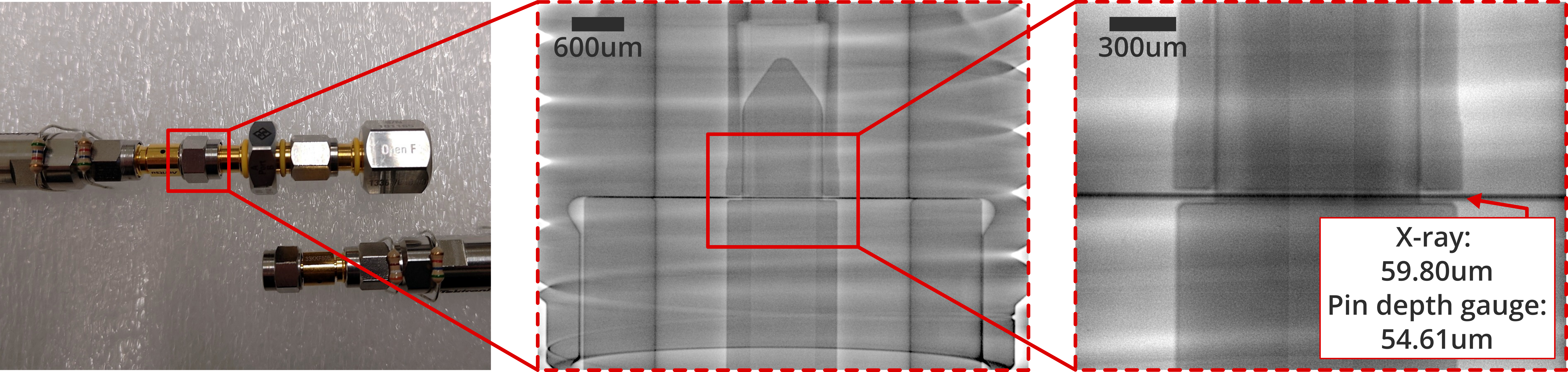}}
	\caption{X-ray inspection of paired standards: (a) load standard, (b) female-female adapter (network), and (c) load connected with a female-male adapter (network-load).}
	\label{fig:5.7xx}
\end{figure}

Table~\ref{tab:5.1} indicates that all one-port standards have the same pin gap at both ports, while the female-female adapter deviates slightly. Furthermore, the table reveals that the adapter standard used to create the network-load standard (female-male) has the largest pin gap at the port interface (i.e., $54.61\,\mu\mathrm{m}$). This discrepancy is also apparent in the X-ray images in Fig.~\ref{fig:5.7xx}.

\begin{figure}[th!]
	\centering
	\includegraphics[width=0.95\linewidth]{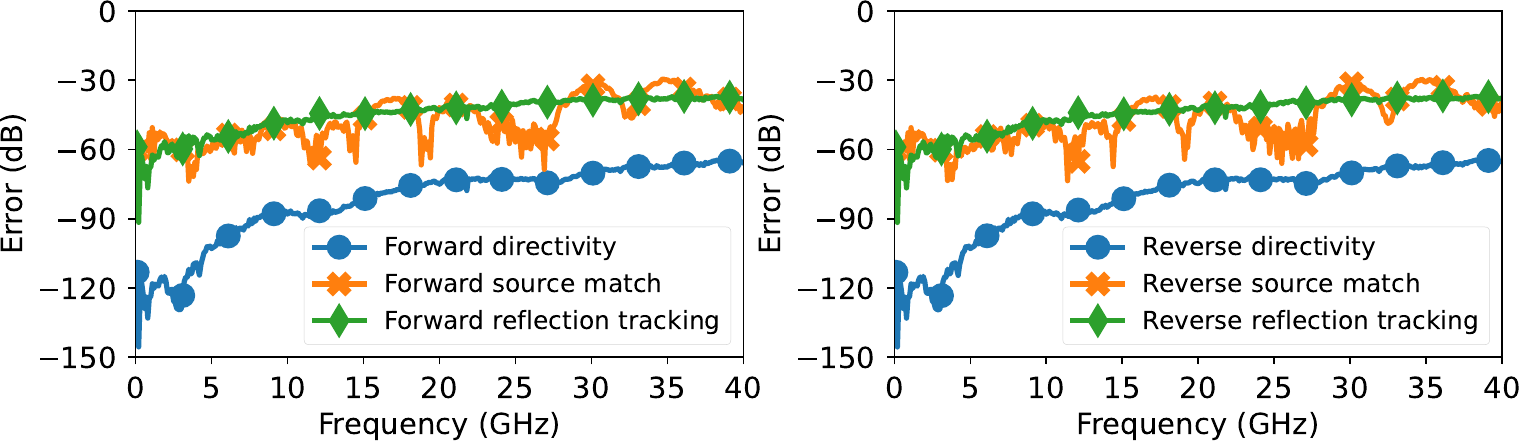}
	\caption{The magnitude of the error vector of the VNA's error terms between SOLR and SRM methods.}
	\label{fig:5.7}
\end{figure}

The ripple observed in the magnitude response of the offset short standard is also noticeable when analyzing the error between the extracted error terms of the SOLR and SRM calibrations, as shown in Fig.~\ref{fig:5.7}. It is clear that both ports have ripple in the source match term. The source match term describes the reflection at the calibration plane, where the effects of the pin gap would be most pronounced \cite{EURAMET2018}. This could also explain the fact that the SOLR calibration did not show such ripple, since all one-port standards have the same pin gap, and the discrepancy in the female-female adapter is not relevant since it is only used to solve the 7th error term, which depends only on the reciprocity property of the network.

\begin{figure}[th!]
	\centering
	\includegraphics[width=0.95\linewidth]{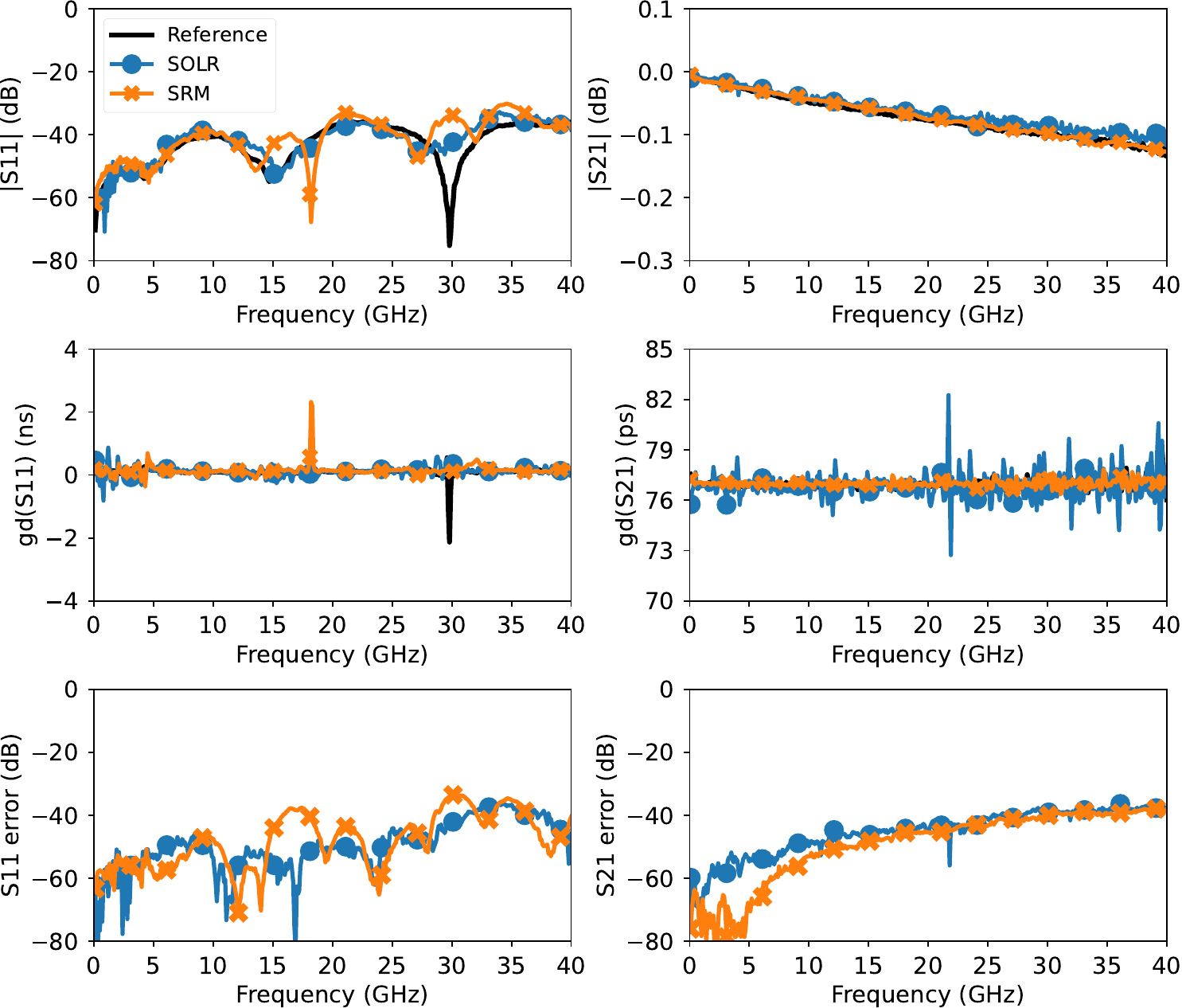}
	\caption{Comparison of the calibrated female-female adapter using SOLR and SRM methods.}
	\label{fig:5.8}
\end{figure}

A final comparison is made comparing the calibrated female-female adapter with both calibration methods. In both SOLR and SRM methods, the adapter was assumed to be unknown but reciprocal during the calibration process. The reference S-parameters of the adapter were provided by the manufacturer and used to establish the error metric. However, no uncertainty bounds were available. Fig.~\ref{fig:5.8} depicts the calibrated adapter derived from both SOLR and SRM methods. These measurement results are compared to the reference S-parameters of the adapter. Both calibration procedures deliver comparable results with similar errors.

Although SOLR and SRM delivered similar results in this experimental example, it is important to note that for the SOLR method, all SOL standards already have been characterized beforehand, whereas for the SRM method only the match standard must be characterized. In addition, it is noteworthy that we achieved results comparable to metrology-level calibration using only S-parameter definition of a single standard, namely the match standard, which sets the reference impedance.This can be particularly advantageous when using economical coaxial calibration kits, as the S-parameters of the open and short standards do not need to be specified.

\subsection{Statistical Numerical Analysis}

\begin{figure}[th!]
	\centering
	\includegraphics[width=0.95\linewidth]{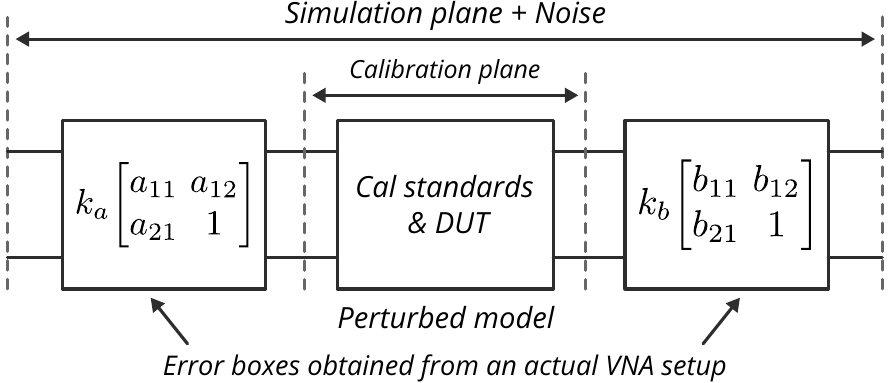}
	\caption{Block diagram illustration of the numerical simulation concept to generate realistic synthetic data for the MC analysis.}
	\label{fig:5.1} 
\end{figure}

The procedure for the numerical analysis involves creating synthetic data of CPW standards using the model developed in \cite{Phung2021a,Schnieder2003,Heinrich1993}. To emulate an on-wafer setup accurately, we utilize error boxes from an actual on-wafer setup that was extracted using multiline TRL calibration on an ISS kit. Details on the measurement setup can be found in \cite{Hatab2023}, where the accuracy of the CPW model was tested, and the measurement datasets are available via \cite{Hatab2023a}. In this numerical setting, the objective is to generate SRM standards based on the CPW model and embed them in the error boxes of the actual VNA setup, introducing different randomness at each iteration to perform the MC analysis. A block diagram summarizing this numerical experiment is depicted in Fig.~\ref{fig:5.1}.

Regarding the geometric parameters of the CPW structure used for simulation, we employed the following values, which are based on the measured ISS \cite{Hatab2023}: signal width of $49.1\,\mu\mathrm{m}$, ground width of $273.3\,\mu\mathrm{m}$, conductor spacing of $25.5\,\mu\mathrm{m}$, and conductor thickness of $4.9\,\mu\mathrm{m}$. The substrate is made of lossless Alumina with a dielectric constant of $9.9$. The conductor is made of gold with conductivity of $41.1\,\mathrm{MS/m}$.

\begin{figure}[th!]
	\centering
	\subfloat[]{\includegraphics[width=.5\linewidth]{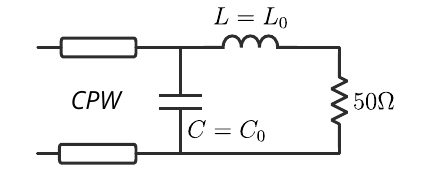}}
	\subfloat[]{\includegraphics[width=.5\linewidth]{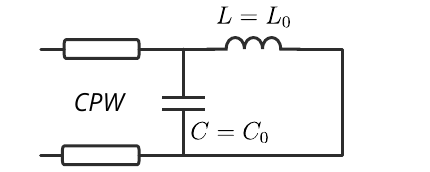}}\\[-1pt]
	\subfloat[]{\includegraphics[width=.5\linewidth]{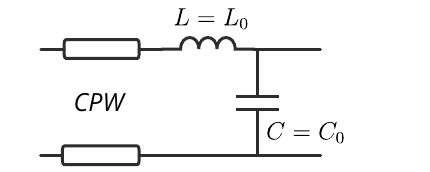}}
	\caption{Models used to simulate non-ideal load standards (a) $50\,\Omega$ match standard with $L_0=5\,\mathrm{pH}, C_0=0.5\,\mathrm{fF}$, (b) short standard with $L_0=10\,\mathrm{pH}, C_0=0.5\,\mathrm{fF}$, and open standard with $C_0=10\,\mathrm{fF}, L_0=0.5\,\mathrm{pH}$. All standards are offset by a $200\,\mu\mathrm{m}$ CPW line segment.}
	\label{fig:5.2}
\end{figure}

For the SRM standards, we implemented match, short, and open standards as non-ideal standards, as shown in Fig.~\ref{fig:5.2}. To create the network-load standards, we used a $4\,\mathrm{mm}$ CPW line as the reciprocal standard, which is combined with the non-ideal match, short, and open standards. Additionally, as discussed in Section~\ref{sec:4}, we created half network-load standards using half of the reciprocal standard, i.e., a $2\,\mathrm{mm}$ CPW line. The reference impedance for both ports was set to $Z_a^{(ref)}=Z_b^{(ref)}=50\,\Omega$. It is worth noting that in the SRM calibration procedure, all standards are not specified, except for the match standard that sets the reference impedance.

\begin{figure}[th!]
	\centering
	\includegraphics[width=0.95\linewidth]{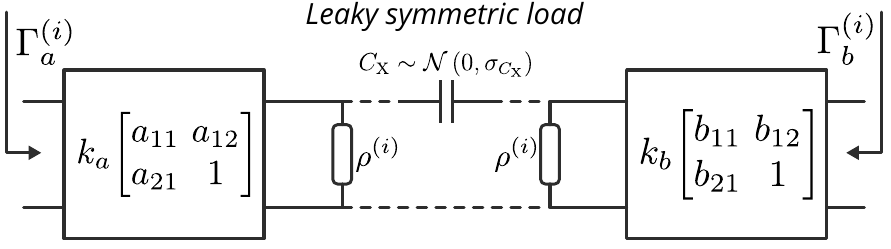}
	\caption{Block diagram showing the inclusion of crosstalk in the MC analysis.}
	\label{fig:5.1x} 
\end{figure}

Various sources of uncertainty were considered in the MC analysis, including VNA noise, asymmetry in the one-port standards, variation in the reciprocal network, variation in the match standard, and crosstalk. To model VNA noise in the MC analysis demonstration, Gaussian noise with a standard deviation of $\sigma_\mathrm{noise}=0.001$ was employed \cite{Hatab2023}. To introduce asymmetry, we introduced a 10\% Gaussian variation in the lumped elements of the one-port standards in Fig.~\ref{fig:5.2} and cross-sectional variation in the CPW offset segment \cite{Hatab2023}. Similary, the reciprocal standard was varied by adjusting the CPW cross-section parameters \cite{Hatab2023} and the length uncertainty of $\pm20\,\mu\mathrm{m}$. The match standard was created separately and perturbed similarly to the one-port standards. To introduce crosstalk, we included a capacitive coupling between the symmetric one-port standards using a randomly assigned capacitor, as shown in Fig.~\ref{fig:5.1x}. The capacitance has a standard deviation of $\sigma_{C_\mathrm{X}}=0.25\,\mathrm{fF}$, corresponding to a standard deviation coupling of approximately $-30\,\mathrm{dB}$ at $150\,\mathrm{GHz}$.

\begin{figure}[th!]
	\centering
	\includegraphics[width=0.95\linewidth]{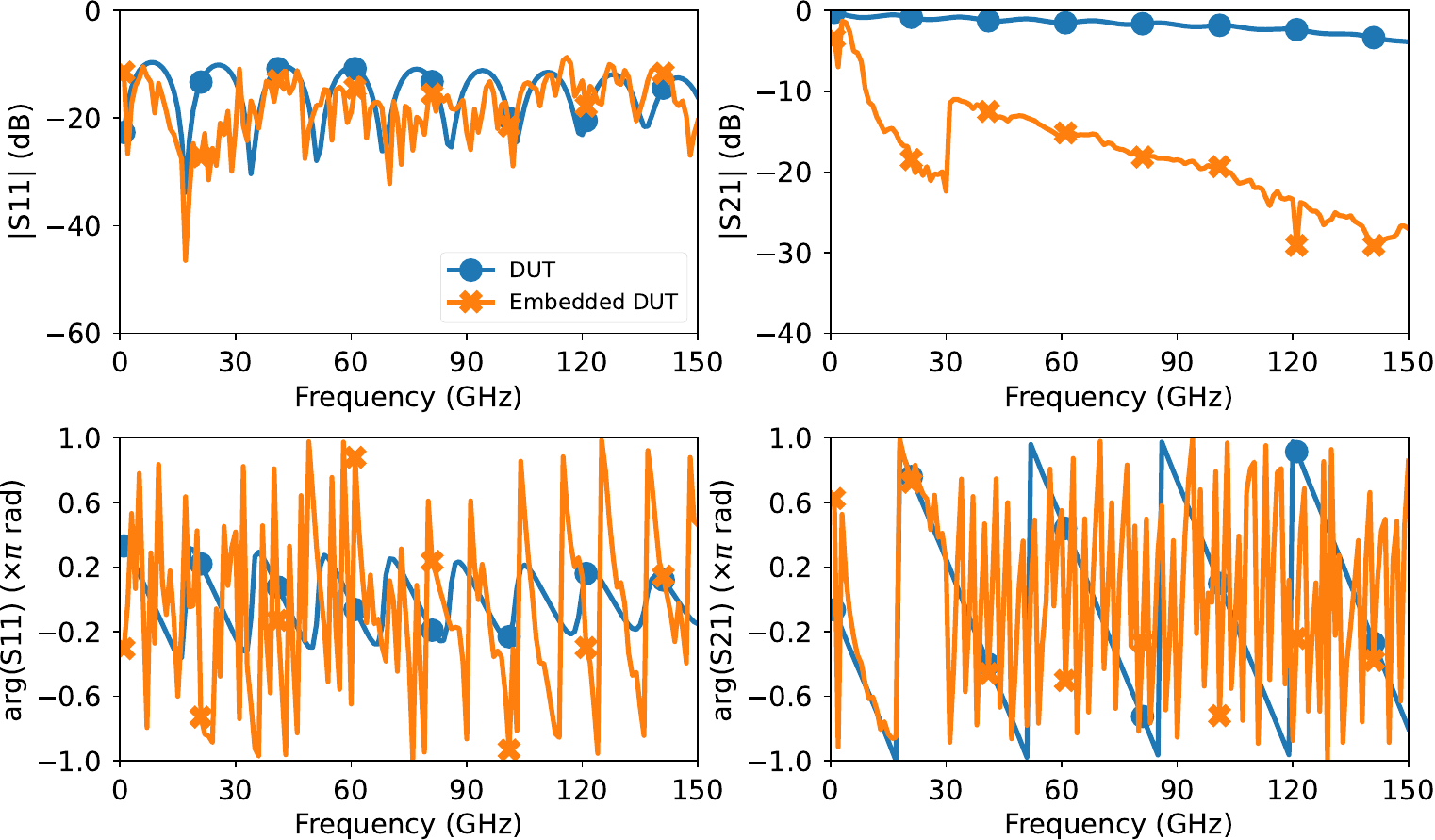}
	\caption{DUT S-parameter response before and after embedding within the error boxes.}
	\label{fig:5.3}
\end{figure}

To verify the accuracy of the calibration, we included a stepped impedance line as DUT, which uses the same CPW structure with the only exception of signal width equal to $15\,\mu\mathrm{m}$. The data has been processed using Python with the help of the package \textit{scikit-rf} \cite{Arsenovic2022}. The frequency response o of the DUT before and after embedding it in the error boxes is depicted in Fig.~\ref{fig:5.3}.

\begin{figure}[th!]
	\centering
	\includegraphics[width=0.95\linewidth]{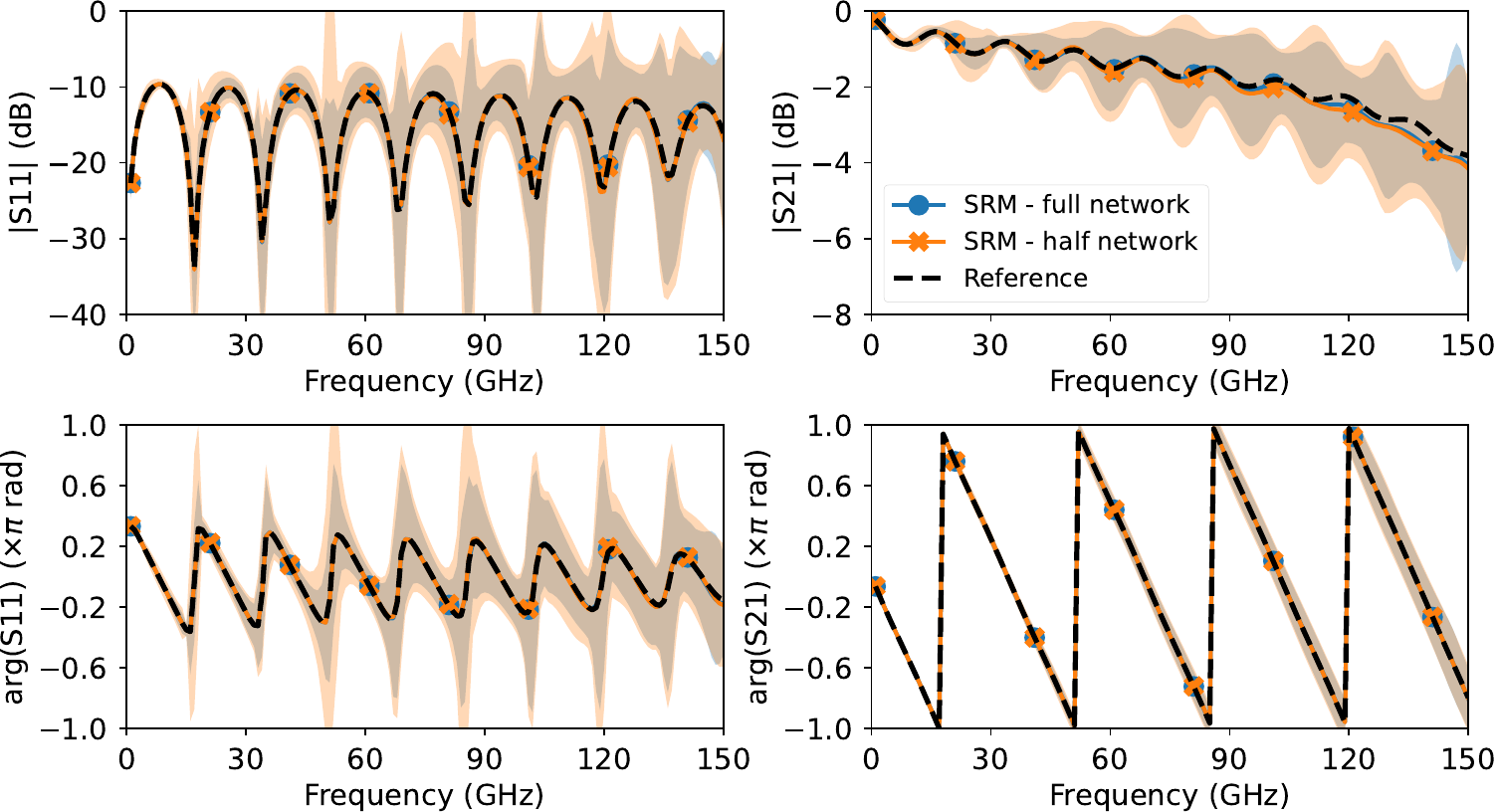}
	\caption{The mean-value and the uncertainty bounds of the calibrated DUT from the MC analysis. The uncertainty bounds are reported as 95\% Gaussian distribution coverage.}
	\label{fig:5.4}
\end{figure}

After conducting 5000 runs of the MC analysis, we obtained the results illustrated in Fig.~\ref{fig:5.4}. The figure displays the mean-value of the calibrated DUT and the uncertainty bounds based on the full and half network variants. As can be seen from the figure, the mean-value of the MC analysis is in agreement with the reference data of the DUT, indicating a proper convergence of the MC simulation. On the other hand, the uncertainty bounds of both full and half network variants show similar values, with the half network showing slightly higher uncertainty, which is probably due to the fact that the reciprocal network used in the network-load standards requires a stricter requirement of being half of the reciprocal network.

\begin{figure}[th!]
	\centering
	\includegraphics[width=0.95\linewidth]{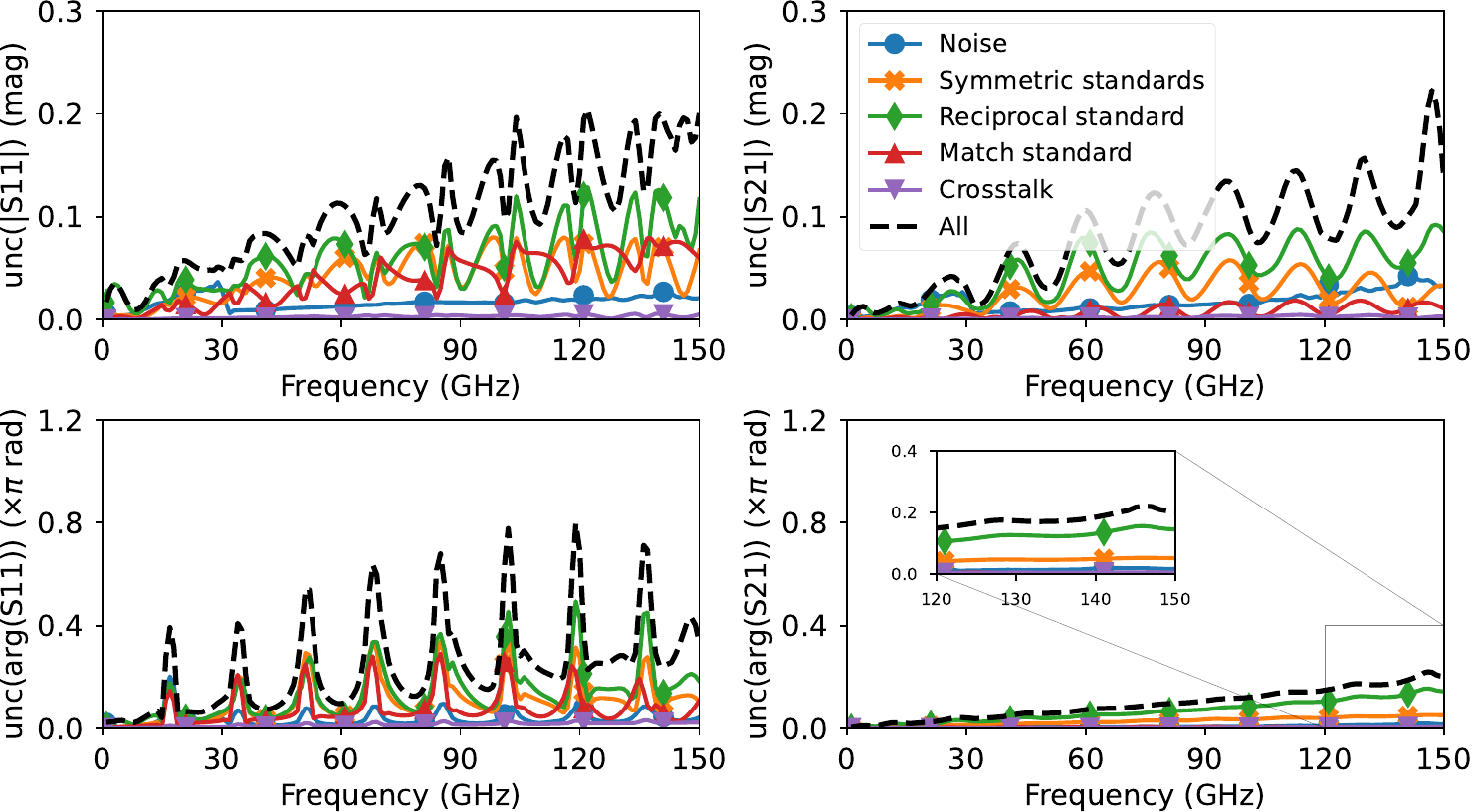}
	\caption{Uncertainty budget of the calibrated DUT due to various uncertainty sources in SRM calibration based on the full network approach, reported as 95\% Gaussian distribution coverage.}
	\label{fig:5.4x}
\end{figure}

\begin{figure}[th!]
	\centering
	\includegraphics[width=0.95\linewidth]{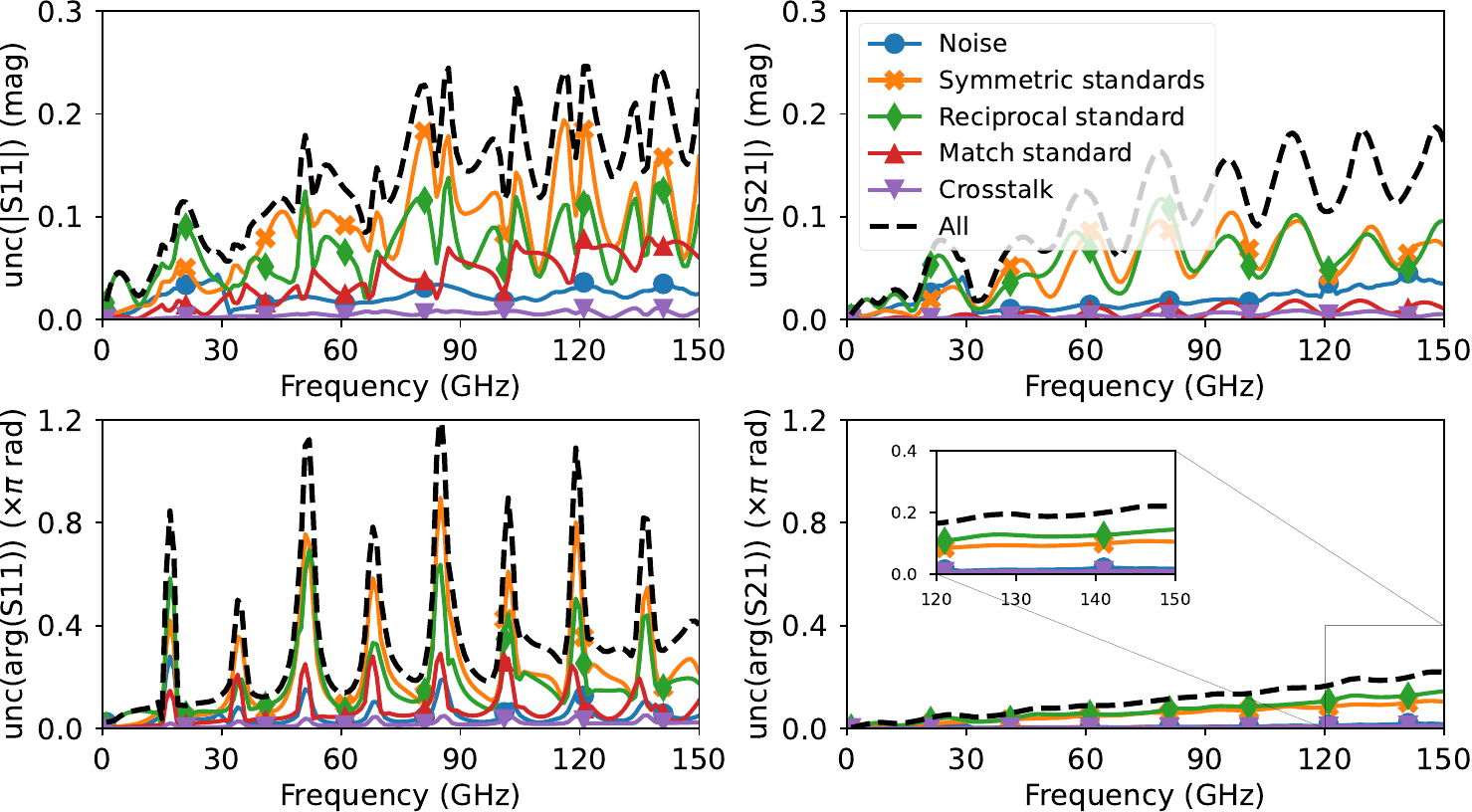}
	\caption{Uncertainty budget of the calibrated DUT due to various uncertainty sources in SRM calibration based on the half network approach, reported as 95\% Gaussian distribution coverage.}
	\label{fig:5.4xx}
\end{figure}

To examine the impact of each uncertainty source, we repeated the MC analysis, considering each uncertainty source individually. The results in Fig.~\ref{fig:5.4x} and Fig.~\ref{fig:5.4xx} show the uncertainty budget for the full and half network cases, respectively. The calibration process is mainly influenced by the symmetric and reciprocal standards since these standards must be symmetric for the one-port standards, and the reciprocal network needs to be replicated in the network-load standards. Interestingly, the smallest contributor to the uncertainty budget is the crosstalk. Noise has a minor effect compared to the uncertainty in calibration standards, but is slightly more significant than crosstalk. Finally, the match standard shows most of its impact in the $S_{11}$ response, while a minor impact can be seen in the $S_{21}$ response.

As already observed in Fig.~\ref{fig:5.4}, the half-network approach shows a slightly higher overall uncertainty than the full-network approach. However, it should be noted that only with the half-network approach is it possible to implement the standards at a constant distance, as illustrated in Fig.~\ref{fig:4.1}.
~

% EOF

\section{Conclusion}
\label{sec:6}

This article presents a new VNA calibration method based on partially defined standards. The proposed SRM method uses one-port symmetric standards, a two-port reciprocal device, a combination of the reciprocal device with the one-port device, and a match standard. Only the match standard must be characterized among all standards, defining the calibration’s reference impedance.

We have extended our proposed method to the particular case of an on-wafer setup, where the probes are fixed in distance. To do this, we restricted the two-port reciprocal device to be symmetric, allowing us to use half of it to define the network-load standards.

To validate the effectiveness of the proposed SRM method, we conducted coaxial measurements, demonstrating its capability to achieve results comparable to metrology-level accuracy while relying solely on the provided S-parameters of the match standard. Additionally, an MC analysis of the SRM method was performed using synthetic data based on CPW model and error boxes derived from an actual on-wafer measurement setup. This numerical analysis aimed to evaluate various sources of uncertainty impacting the calibration process. Our findings highlighted the significance of variations in symmetrical standards and the influence of the reciprocal network, which plays a crucial role as part of the network-load standards. Interestingly, crosstalk showed minor influence compared to other sources of uncertainties.

%\appendices
%\input{Sections/AppendixA}

\section*{Acknowledgment}
The authors thank the ebsCENTER, Graz, Austria for providing access to their X-ray equipment.

\bibliographystyle{IEEEtran}
\bibliography{References/references.bib}

\begin{IEEEbiography}[{\includegraphics[width=1in,height=1.25in,clip,keepaspectratio]{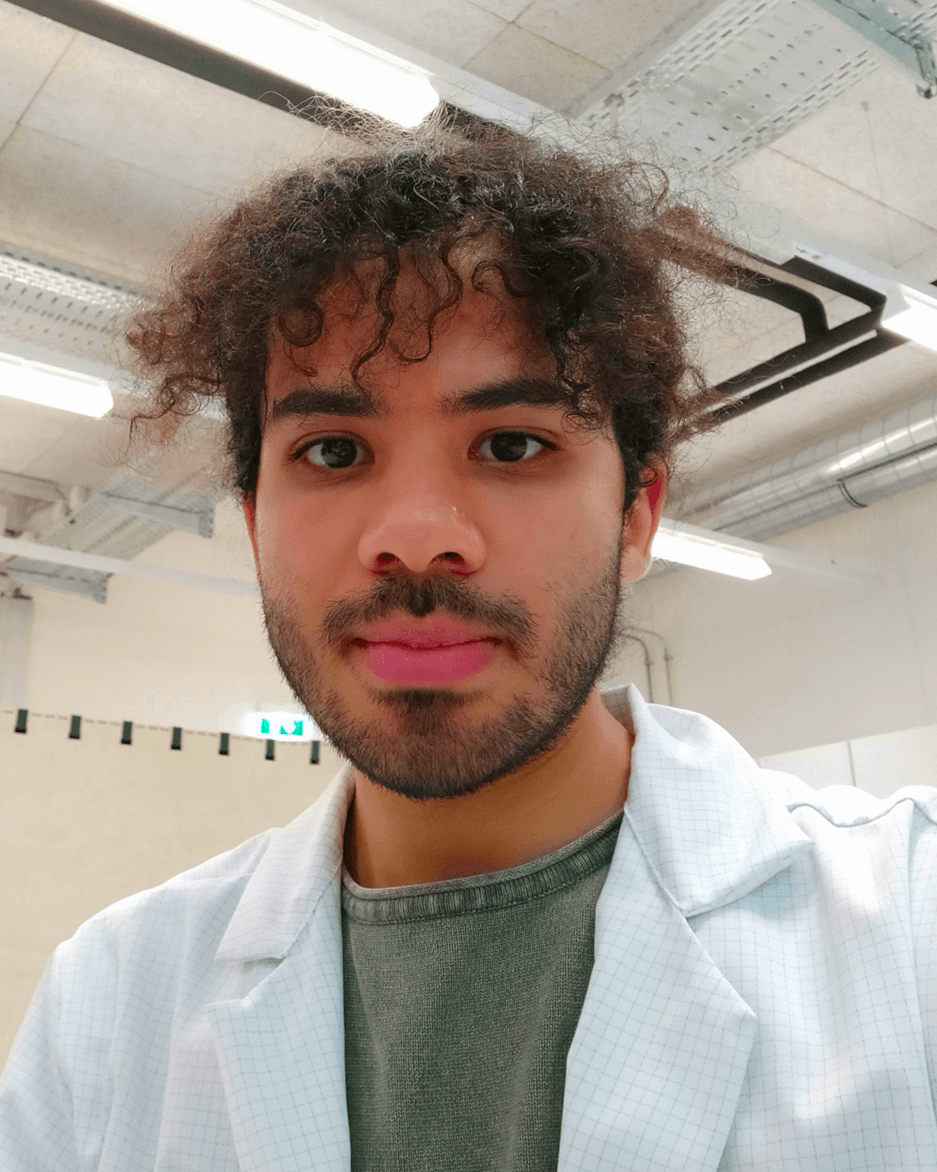}}]{Ziad Hatab } (Student Member, IEEE) received the B.Sc. and Dipl.-Ing.(M.Sc.) degrees in electrical engineering from the Graz University of Technology, Graz, Austria, in 2018 and 2020, respectively, where he is currently pursuing the Ph.D. degree with the Institute of Microwave and Photonic Engineering. He joined the Christian Doppler Laboratory for Technology Guided Electronic Component Design and Characterization (TONI), Graz University of Technology, as a Research Member, in 2020. His research focuses on measurement techniques and calibration methods at millimeter-wave frequencies and beyond.
\end{IEEEbiography}

\begin{IEEEbiography}[{\includegraphics[width=1in,height=1.25in,clip,keepaspectratio]{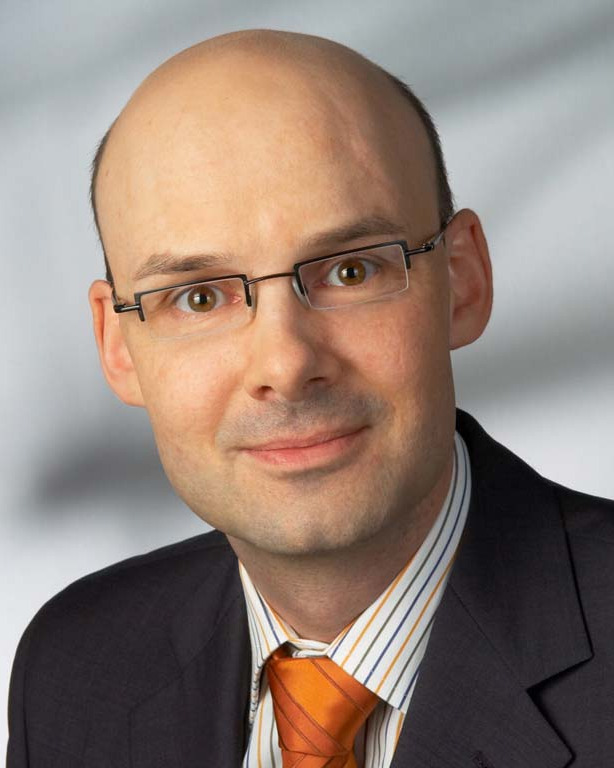}}]{Michael Ernst Gadringer } (Senior Member, IEEE) is an Associate Professor at the Institute of Microwave and Photonic Engineering at Graz University of Technology, Austria. He received the Dipl.-Ing. and Dr.techn. degrees from Vienna University of Technology, Austria, in 2002 and 2012, respectively. During his studies, he was involved in designing analog and digital linearization systems for power amplifiers and behavioral modeling of microwave circuits. His current research focuses on planning and implementing complex measurements, emphasizing calibration techniques and material characterization. Michael Gadringer has authored more than 20 journals and 50 conference papers. He was a member of the IEEE 1765 standard working group on the recommended practice for estimating the Error Vector Magnitude of digitally modulated signals. In addition, he is contributing to the IEEE P2822 working group on the recommended practice for Microwave, Millimeter-wave, and THz on-wafer calibrations, deembedding, and measurements.
\end{IEEEbiography}
\begin{IEEEbiography}[{\includegraphics[width=1in,height=1.25in,clip,keepaspectratio]{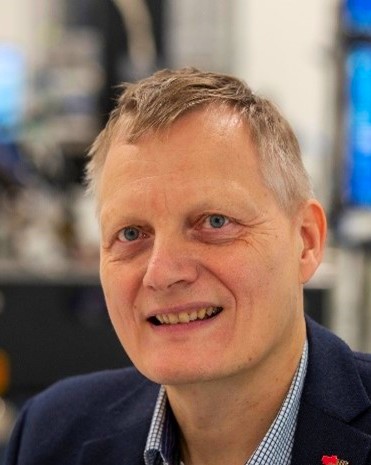}}]{Wolfgang Bösch }
(Fellow, IEEE) received the Dipl.Ing. degree from the Technical University of Vienna, Vienna, Austria, in 1985, the Ph.D. degree from the Graz University of Technology, Graz, Austria, in 1988, and the M.B.A. degree from the School of Management, University of Bradford, Bradford, U.K., in 2004. In 2010, he joined the Graz University of Technology to establish the Institute for Microwave and Photonic Engineering. For the last eight years, he was also the Dean of the Faculty of Electrical and Information Engineering, which currently incorporates 13 institutes and 20 full professors covering the areas of energy generation and distribution, electronics, and information engineering. He is responsible for the strategic development, budget, and personnel of the faculty. Prior to this, he was the Chief Technology Officer (CTO) of the Advanced Digital Institute, Shipley, U.K. He was also the Director of Business and Technology Integration with RFMD, Newton Aycliffe, U.K. For almost ten years, he was with Filtronic plc, Leeds, U.K., as the CTO of Filtronic Integrated Products and the Director of the Global Technology Group. Before joining Filtronic, he held positions at the European Space Agency, Noordwijk, The Netherlands, working on amplifier linearization techniques; MPR-Teltech, Burnaby, BC, Canada, working on MMIC technology projects; and the Corporate Research and Development Group, M/A-COM, Boston, MA, USA, where he worked on advanced topologies for high-efficiency power amplifiers. For four years, he was with DaimlerChrysler Aerospace (currently, Hensoldt), Ulm, Germany, working on T/R modules for airborne radar. He has published more than 180 articles and holds four patents. Prof. Bösch is a Fellow of IET.
\end{IEEEbiography}

\vfill

\end{document}